\documentclass[12pt]{article}

\usepackage{graphics}
\usepackage{color}
\usepackage{amssymb}
\usepackage{latexsym}

\font\manual=manfnt \def\dbend{\lower3.5pt\hbox{\manual\char127}}

\def\ie{{\it i.e.}}
\def\eg{{\it e.g.}}
\def\cf{{\it c.f.}}

\def\sst{\scriptscriptstyle}

\def\frac#1#2{{#1\over#2}}
\def\coeff#1#2{{\textstyle{#1\over #2}}}
\def\half{\frac12}
\def\hf{{\textstyle\half}}

\def\IR{{\mathbb R}}
\def\IC{{\mathbb C}}

\def\IP{{\mathbb P}}
\def\IZ{{\mathbb Z}}

\catcode`\@=11
\def\slash#1{\mathord{\mathpalette\c@ncel{#1}}}
\def\underrel#1\over#2{\mathrel{\mathop{\kern\z@#1}\limits_{#2}}}

\catcode`\@=12
\def\ket#1{|#1\rangle}

\def\vev#1{\langle#1\rangle}

\def\sinh{{\rm sinh}} 	
\def\cosh{{\rm cosh}}

\def\exp{{\rm exp}}

\def\NN{{\cal N}}

 \def\CS{{\cal S}}
\def\TT{{\cal T}} 
 
\def\VV{{\cal V}} 
\def\WW{{\cal W}}

\def\unlockat{\catcode`\@=11}
\def\lockat{\catcode`\@=12}
\unlockat
\def\newsec#1{\global\advance\secno by1\message{(\the\secno. #1)}
\global\subsecno=0\global\subsubsecno=0\eqnres@t\noindent
{\bf\the\secno. #1}
\writetoca{{\secsym} {#1}}\par\nobreak\medskip\nobreak}
\global\newcount\subsecno \global\subsecno=0
\def\subsec#1{\global\advance\subsecno
by1\message{(\secsym\the\subsecno. #1)}
\ifnum\lastpenalty>9000\else\bigbreak\fi\global\subsubsecno=0
\noindent{\it\secsym\the\subsecno. #1}
\writetoca{\string\quad {\secsym\the\subsecno.} {#1}}
\par\nobreak\medskip\nobreak}
\global\newcount\subsubsecno \global\subsubsecno=0
\def\subsubsec#1{\global\advance\subsubsecno by1
\message{(\secsym\the\subsecno.\the\subsubsecno. #1)}
\ifnum\lastpenalty>9000\else\bigbreak\fi
\noindent\quad{\secsym\the\subsecno.\the\subsubsecno.}{#1}
\writetoca{\string\qquad{\secsym\the\subsecno.\the\subsubsecno.}{#1}}
\par\nobreak\medskip\nobreak}
\def\subsubseclab#1{\DefWarn#1\xdef
#1{\noexpand\hyperref{}{subsubsection}%
{\secsym\the\subsecno.\the\subsubsecno}%
{\secsym\the\subsecno.\the\subsubsecno}}%
\writedef{#1\leftbracket#1}\wrlabeL{#1=#1}}
\lockat
\newcommand{\be}{\begin{equation}}
\newcommand{\ee}{\end{equation}}

\newcommand{\bbb}{\begin{eqnarray}}
\newcommand{\eee}{\end{eqnarray}}
\newcommand{\pref}[1]{(\ref{#1})}

\begin{document}
 

%

\def\aa{{\bf a}}
\def\ads{AdS}
\def\btz{{\sst BTZ}}
\def\mbar{{\bar m}}
\def\N{{\bf N}}
\def\P{{\bf P}}
\def\S{{S}}
\def\T{{T}}
\def\pp{{\bf p}}
\def\str{{\sst\rm str}}
\def\lstr{\ell_{s}}
\def\lpl{\ell_{p}}
\def\rads{\ell}

%
\def\Lpotl{1}
\def\dsdomains{2}
\def\openstr{3}
\def\topinf{4}
\def\fractal{5}   

\begin{titlepage}
\rightline{EFI-03-10}

\rightline{hep-th/0303087}

\vskip 3cm
\centerline{\Large{\bf Closed String Tachyon Condensation}}
\medskip
\centerline{\Large{\bf and Worldsheet Inflation}}

\vskip 2cm
\centerline{
Bruno Carneiro da Cunha\footnote{\texttt{bcunha@theory.uchicago.edu}}%
~~ and ~~%
Emil J. Martinec\footnote{\texttt{e-martinec@uchicago.edu}}}
\vskip 12pt
\centerline{\sl Enrico Fermi Inst. and Dept. of Physics}
\centerline{\sl University of Chicago}
\centerline{\sl 5640 S. Ellis Ave., Chicago, IL 60637, USA}

\vskip 2cm

\begin{abstract}
Closed string tachyon condensation in spacetime generates
potentials on the worldsheet that model two-dimensional
inflationary cosmology.  These models illustrate and
elucidate a variety of aspects of inflation, in particular
the generation of quantum fluctuations and their
back-reaction on geometry.
We exhibit a class of Liouville gravity models
coupled to matter that can exhibit, for example:
(a) pure de~Sitter gravity;
(b) slow-roll inflation; (c) topological inflation;
and (d) graceful exit into an FRW phase.  
The models also provide a quantitative testing ground 
for ideas about the origin of inflation,
such as the various `no-boundary/tunnelling' proposals,
and the `eternal/chaotic' inflationary scenario.
We suggest an alternative mechanism for quantum
creation of cosmological spacetimes which, 
in the context of the model, provides a natural
explanation for why the typical FRW cosmology
at large scales underwent a period of inflation
at small scale.
\end{abstract}

\end{titlepage}

\newpage

\setcounter{page}{1}


\section{\label{introsec}Introduction}

The development of precision cosmology 
(\cf\ \cite{Sarkar:2002gc,Turner:2002fh,Turner:2002ts}
for recent reviews) has provided a stunning confirmation of 
the basic predictions of inflationary theory
(\cf\ \cite{Lyth:1998xn,Liddle:2000cg} for reviews).
Nevertheless, many questions yet remain:
What is the origin of inflation?  What is the inflaton?
Is the standard linearized analysis of fluctuations correct?
Is inflation a generic phenomenon?  
Why is the current value of the cosmological constant so small?
Are there reasonable alternative explanations of the data?
There is much still to be understood.

Insight can often be gained by the development of simple
models that capture key features of the phenomenon under study.
Two-dimensional quantum field theory models have
a rich history of mirroring key aspects of their
higher-dimensional siblings; for example,
dimensional transmutation (the $\IC\IP^n$ model), 
spontaneous symmetry breaking (the Gross-Neveu model),
confinement (the Schwinger and 't~Hooft models), 
anomalies and topology (the Wess-Zumino-Witten model
among others), instanton effects (many examples), and
supersymmetry (many more examples), to name but a few.

Two-dimensional models have also been a fruitful laboratory 
for the investigation of 
quantum gravity and its interaction with matter, 
for example in the context of string theory
in two or less target space dimensions
(\cf\ \cite{Ginsparg:1993is} for a review);
and in the investigation of quantum black holes
(\cf\ \cite{Harvey:1992xk,Thorlacius:1995ip,Martinec:1996ad} for reviews).
Our interest here lies in the development of two-dimensional
models which may shed light on some of the issues
surrounding inflationary cosmology mentioned above.

As we review below in section \ref{liourev},
two-dimensional (anti) de~Sitter cosmology 
is the Liouville gravity model.
It arises as the worldsheet description of a test string propagating
in the presence of a homogeneous closed string tachyon condensate,
in string theory above the critical dimension
\cite{Polchinski:1989fn,Cooper:1991vg,Cooper:1991zc}.
In this setting,
the target space time coordinate has an interpretation
as the scale factor of the $2d$ worldsheet metric
(the Liouville field $\phi$), with
the spatial coordinates of the target playing the
role of (conformal) matter fields.

Interesting models of inflation arise upon coupling
the gravity and matter degrees of freedom.
The example we focus on primarily is the Liouville-Sine-Gordon
theory, where the matter potential is $\cos(kX)$
for some scalar field $X$.  When gravitationally
dressed, this potential describes an inhomogenous
condensate of the closed string tachyon.
For sufficiently small $k$, this model satisfies
the criteria of slow-roll inflation.  For a particular
value of $k$, the model turns out to be exactly solvable --
it is simply the complexification of Liouville theory.
We exhibit a number of elementary solutions, describing universes
that inflate, reach a maximum scale size,
and then recollapse.   We also discuss
an amusing example of topological (domain wall) inflation.

A shortcoming of these models is that the target space
background is not fully believable as a consistent string
background; the tachyon condensate that is
responsible for eternal worldsheet de~Sitter space
cannot persist indefinitely,
rather it must saturate at some point and allow
the target space geometry to back-react.
This would lead to a transition to some other phase of
the worldsheet cosmology; however, we do not have sufficient
understanding of the target space dynamics to determine
what worldsheet dynamics (if any) emerges.
However, condensation of a closed string tachyon localized
on a defect does not suffer this problem -- 
there is strong evidence that the tachyon 
eventually relaxes to a smooth time-dependent
solution of the low-energy string equations
\cite{Adams:2001sv,Vafa:2001ra,Harvey:2001wm,David:2001vm}.
Models of this sort provide an intriguing analogue
of our universe, exhibiting an early inflationary
phase followed by relaxation of the cosmological
constant into an FRW phase.  We give an example of
such a model in section \ref{localtach}.

We then begin, in section \ref{Qeff},
a study of quantum effects in these models,
calculating the quantum stress tensors of matter
as well as the minisuperspace wavefunctions of some
of the solutions.  We also compare several
approaches to $2d$ de~Sitter thermodynamics.

One of our motivations for the construction of these
two-dimensional models was to provide a concrete testing ground
for various scenarios regarding the origin
of inflation, such as the `no-boundary' or `tunnelling' proposals
\cite{%
Hawking:1982fz,%
Hartle:1983ai,%
Vilenkin:1983xq,%
Linde:1984mx,%
Vilenkin:1986cy,%
Vilenkin:1988kf,%
Hawking:1998bn%
},
as well as `eternal' or `chaotic' inflation
\cite{Linde:1983gd,Vilenkin:1983xq,Steinhardt:1982kg,%
Linde:1986fd,Linde:1994xx}.
We comment in section \ref{discsec}
on what our models might have to say about
these proposals, and add a proposal of our own.
The implicit equivalence of the worldsheet metric scale factor
and the target time coordinate in string theory means that
worldsheet inflation is a region of strong scale/time dependence;
quantum processes such as string (universal) creation in this
time-dependent background generate an ensemble of $2d$ universes
where none were present before,
and thus a probability measure on cosmologies
at late times/large scales.

We thus anoint ourselves as Gedankeng\"otter
of the two-dimensional cosmos, and proceed to 
create the universe.


\section{\label{liourev}Review of Liouville theory}

The ordinary bosonic Liouville action is
(\cf\ \cite{Ginsparg:1993is}, whose conventions we follow)
\be\label{Sliou}
  \CS_{\sst\rm Liouville} = 
	\frac1{4\pi}\int \! d\tau d\sigma\,\sqrt{-\hat g}\Bigl(
	-\hf(\hat\nabla\varphi)^2-\hf{Q}R(\hat g)\varphi
	-\frac{\mu}{2\gamma^2}\,e^{\gamma\varphi}\Bigr)\ .
\ee
The field $\varphi$ is to be thought of as the conformal factor
for the 2d metric 
\be\label{twodmetric}
  ds^2=e^{\gamma\varphi}(-d\tau^2+d\sigma^2)\ .
\ee
In the classical theory, $Q=2/\gamma$,
and the classical equations of motion describe metrics
of constant curvature
\be\label{constcurv}
  R(e^{\gamma\varphi}\hat g)=-\frac{\mu}{2}\ .
\ee
Thus one can interpret $\mu$ as (minus) the cosmological constant.
The general classical solution for $\varphi$ can be expressed locally as
\be\label{gensoln}
  e^{\gamma\varphi} = -\frac{16}{\mu}
	\frac{\partial A(x^+)\partial B(x^-)}{[A(x^+)-B(x^-)]^2}\ ,
\ee
where $x^\pm=\tau\pm\sigma$.  For $\mu>0$, one has
two-dimensional AdS spacetime; the only spatially homogeneous
solution is
\be\label{adssoln}
  e^{\gamma\varphi} = \frac{4}{\mu}\,
	\frac{\varepsilon^2}{\cosh^2(\varepsilon \tau)}\ ,
\ee
describing an expansion out of a big bang and into a big crunch
singularity.
The usual static AdS geometries 
\bbb
   e^{\gamma\varphi} 
   &=& \frac{4}{\mu}\,\frac{\varepsilon^2}{\sin^2(\varepsilon \sigma)}
	\label{globalads}\\
   &=& \frac{4}{\mu}\,\frac{1}{\sigma^2}
	\label{poincare}\\
   &=& \frac{4}{\mu}\,\frac{\epsilon^2}{\sinh^2(\epsilon \sigma)}\ .
	\label{rindlerads}
\eee
(corresponding to global, Poincar\'e, and Rindler
slices), with a timelike boundary at spatial infinity, 
are not compatible with the compact spatial topology we want to impose 
to make contact with string theory;%
\footnote{Rather, they are compatible with a timelike identification,
a fact we will make use of below.}
thus we find only a patch of global AdS for spacetimes
of cylindrical topology.

For $\mu<0$, there are three spatially homogeneous solutions:
\bbb
   e^{\gamma\varphi} &=& \frac{4}{-\mu}\,
	\frac{\varepsilon^2}{\sinh^2(\varepsilon \tau)}
	\label{milneds}\\
   &=& \frac{4}{-\mu}\,\frac{1}{\tau^2}
	\label{flatds}\\
   &=& \frac{4}{-\mu}\,\frac{\epsilon^2}{\sin^2(\epsilon \tau)}\ .
	\label{globalds}
\eee
Evaluating the stress-energy tensor
\bbb
  T_{++} &=& \hf(\partial_+\varphi)^2-\hf Q\partial_+^2\varphi
\nonumber\\
  T_{--} &=& \hf(\partial_-\varphi)^2-\hf Q\partial_-^2\varphi
\label{stresstens}
\eee
of these solutions,
one finds the energy $E_L-\frac18 Q^2=\varepsilon^2/2\gamma^2>0$ 
for the first, `hyperbolic' solution \pref{milneds}; 
$E_L-\frac18 Q^2=0$ for the second, `parabolic' solution \pref{flatds}; 
and $E_L-\frac18 Q^2=-{\epsilon^2}/{2\gamma^2}<0$ 
for the third, `elliptic' solution \pref{globalds}, 
see figure \Lpotl.

\begin{figure}[ht]
\begin{center}
\[
\mbox{\begin{picture}(240,136)(0,0)
\includegraphics{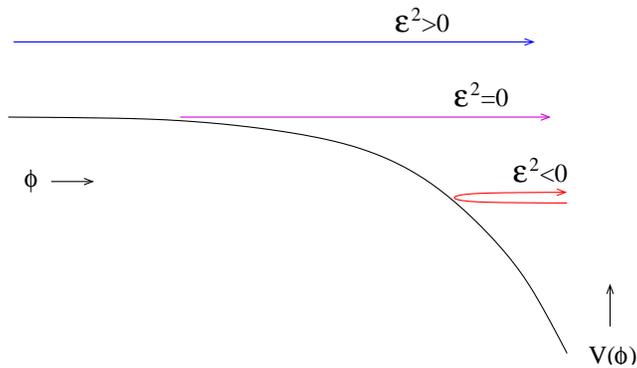}
\end{picture}}
\]
\caption{\it The three classes of solution of de~Sitter Liouville theory,
corresponding to positive, zero, and negative energy
(relative to the Casimir energy on the cylinder $-\coeff18 Q^2$).
Large positive $\phi$ corresponds to large scale factor.
The positive energy solution describes gravity coupled to 
matter energy density; it has an FRW-like `big bang' in the past 
and dS asymptotics in the far future.
The zero energy solution describes de~Sitter geometry in
flat coordinates, while the negative energy solution describes
the `bounce' geometry of global de~Sitter space.
}
\end{center}
\end{figure}

To interpret these solutions
as two-dimensional cosmologies, it is 
useful to transform the conformal time $\tau$ to
proper time $t$ via
$dt = {\sst\sqrt{-\frac{\mu}{4}}}\,e^{\gamma\varphi/2} d\tau$;
in these coordinates the geometries become
\bbb
  ds^2 &=& \frac{4}{-\mu}\left(-dt^2 
		+ \varepsilon^2\,\sinh^2 t\,d\sigma^2\right)
\nonumber\\
	&=& \frac{4}{-\mu}\left(-dt^2 + e^{2t} d\sigma^2\right)
\label{FRW}\\
	&=& \frac{4}{-\mu}\left(-dt^2 
		+ \epsilon^2\,\cosh^2 t\,d\sigma^2\right)\ .
\nonumber
\eee
The first cosmology describes a universe which has a 
past Milne-type singularity at $t=0$ ($\tau=-\infty$)
and expands into an asymptotically de~Sitter future
as $t\to\infty$ ($\tau\to 0$);
the second cosmology describes eternal de~Sitter space in `flat'
coordinates, while the third describes eternal de~Sitter space
in `global' coordinates.  Note that, because we are
identifying $\sigma\sim\sigma+2\pi$, the latter two metrics
are not related by analytic continuation.%
\footnote{Of course, if $\sigma$ were a noncompact coordinate,
the surface $t=-\infty$ in the flat coordinates is just
a past horizon; continuing past it reveals a second flat
coordinate patch needed to cover the entire global de~Sitter space.}
The solutions to the equation of motion
\pref{constcurv} are always locally constant curvature
geometries, and so must cover a portion of the global
conformal diagram of (anti)de~Sitter space; the
relevant domains for the three metrics 
\pref{milneds}, \pref{flatds}, \pref{globalds}
are shown in figure \dsdomains.

\begin{figure}[ht]
\begin{center}
\[
\mbox{\begin{picture}(109,109)(0,0)
\includegraphics{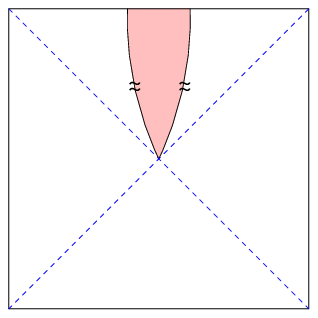}
\end{picture}
\begin{picture}(109,109)(0,0)
\includegraphics{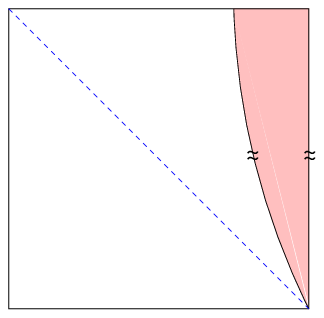}
\end{picture}
\begin{picture}(109,109)(0,0)
\includegraphics{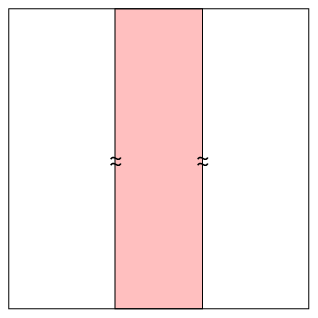}
\end{picture}
}
\]
\caption{\it
Domains of the de~Sitter conformal diagram covered by the metrics 
\pref{milneds}, \pref{flatds}, \pref{globalds}, respectively.
For simplicity, only half of the conformal diagram is shown in each case
(\ie\ the zero-dimensional `angular sphere' consists of two points).
}
\end{center}
\end{figure}


\subsection{\label{STinterp}The spacetime interpretation}

Quantum consistency of two-dimensional gravity requires
cancellation of the conformal anomaly
\be\label{confanom}
  c_{\rm tot} = 1+3Q^2 + c_{\rm matter} - 26 = 0\ ,
\ee
as well as the quantum scale invariance of the exponential
interaction, which amounts to the relation
\be\label{gqrel}
  Q=\frac{2}{\gamma}+\gamma\ ,
\ee
a slight modification of the classical value $Q=2/\gamma$.
The semiclassical limit of the Liouville
theory is the large $c_{\rm matter}$ limit; 
rescaling $\varphi\to\frac{1}{\gamma}\varphi$
in \pref{Sliou}, one sees that $\gamma$ is the coupling
constant of the theory.  There are actually two such limits.
First, one can take $c_{\rm matter}\to-\infty$, 
in which case the Liouville field
is a standard positive metric scalar field. 
Alternatively, one can take $c_{\rm matter}\to\infty$, 
but then \pref{confanom} requires that $Q$ is pure
imaginary; to maintain reality of the action, one
Wick rotates $\varphi=i\phi$ so that $\phi$ is a timelike
(negative metric) scalar field.
We also remove the factors of $i$ 
in the definition of $Q$ and $\gamma$, 
so that for instance $Q=\sqrt{(d-25)/3}$.
Note also that the Liouville energy \pref{stresstens} changes sign.

We will be interested primarily in matter systems with
a string theoretic interpretation as target space geometries,
for which $c_{\rm matter}$ is the spatial dimension $d$.  
These theories were considered as cosmological models
quite some time ago
\cite{Polchinski:1989fn,Cooper:1991vg,Cooper:1991zc}.
We thus take the matter system to be $d$ free fields $Y_i$, $i=1,...,d$;
then the spacetime interpretation of the combined
Liouville plus matter system is that of a test string
in the background of a spatially homogeneous closed string tachyon
condensate in the bosonic string, in the supercritical
spacetime dimension $d+1>26$.  A central theme of our
investigation surrounds the dual interpretations
(introduced in the works just mentioned)
of timelike Liouville theory as describing on the one hand
propagation in target spacetime
of a test string in the presence of a closed string
tachyon condensate, and on the other hand as describing a
de~Sitter phase of the worldsheet geometry.%
\footnote{This idea has resurfaced more recently in
the context of brane worlds, via `mirage' cosmologies:
\cite{Kehagias:1999vr,Papantonopoulos:2000yz,%
Papantonopoulos:2000xs,Kachru:2002kx}, to name a few.}

In the supercritical dimension $d>25$, the dilaton
\be\label{lindil}
  g_{\rm str} = e^D = e^{-Q\phi/2}
\ee
is such that strings are strongly coupled as $\phi\to-\infty$,
which corresponds to small scale factor in the 2d cosmology
\cite{Polchinski:1989fn,Cooper:1991vg,Cooper:1991zc}.
Thus we do not really have consistent control
over the `big bang' region of the two-dimensional theory --
topological fluctuations of the worldsheet are unsuppressed.
Passing to the Einstein frame in spacetime
\be
  (ds_E^{\sst\rm(ST)})^2 = e^{-\frac{4D}{d-1}} 
	(ds_{\rm str}^{\sst\rm(ST)})^2 \ ,
\label{Eframe}
\ee
one sees that the source of the problem is that
the spacetime geometry is itself 
a big bang cosmology
\cite{Antoniadis:1988aa,Craps:2002ii}
\be\label{STmilne}
  (ds_E^{\sst\rm(ST)})^2 =
	-dT^2+\bigl(\coeff{Q}{d-1}\bigr)^2\,T^2dY_i^2
\ee
(where $T=\coeff{d-1}{\sqrt2\,Q}\,e^{\frac{Q\phi}{d-1}}$),
so that early time $\phi\to-\infty$
corresponds to a big bang in the
target space.  Perhaps this problem could be avoided
by having the target space cosmology emerge from
an inflationary phase of the sort described in
\cite{Maloney:2002rr}, but this is a higher order speculation.

Even so, it is not clear to us what pathology in 
the two-dimensional physics ensues if
by hand we restrict the worldsheet topology to be a cylinder.
Alternatively, we could work in the critical dimension $d=25$
where the dilaton can be kept constant and small at early times
(although the $2d$ gravity theory is strongly coupled in this case).
Therefore let us forge ahead with the worldsheet dynamics.
The BRST constraints demand 
a vanishing expectation value of the total worldsheet
stress tensor.  At the semi-classical level, 
the zero mode of this condition is
\be\label{enconstraint}
  E_L+E_{\rm matter}-1
	=-\frac{\varepsilon^2}{2\gamma^2}+E_{\rm matter}-\coeff18{Q^2}-1
	=0\ .
\ee
Note in particular that 
for standard positive energy matter, there is only a finite
range of matter energies for which the global
de~Sitter solution $-\varepsilon^2=\epsilon^2>0$ is allowed,
although that range is large for large $d$ (where $Q^2\sim d/3$).
Since both ends $\tau\to\pm\infty$ of the cosmology \pref{globalds}
correspond to $\phi=+\infty$, these solutions represent
classical production on-shell of pairs of strings 
(which are thus themselves tachyonic)
in the time-dependent tachyon field.%
\footnote{Quantum mechanically, there is an nonzero amplitude
to produce any string state, however for high mass strings 
this proceeds by a tunnelling process and is exponentially 
suppressed.  See for instance \cite{Strominger:2002pc}
and section \ref{Qeff} below.}
The string theory interpretation gives a completely
different perspective on 
the global de~Sitter solution -- one does not view it as
a contracting phase followed by a bounce into
an expanding phase, rather there is a pair of
expanding universes produced at a time 
$\phi\sim \coeff1\gamma\log[\coeff{4\epsilon^2}{-\mu}]$.

Similarly, the solution \pref{flatds} is the dressing
of a string state that is effectively `massless'
in the tachyon background, while the solution \pref{milneds}
dresses a `massive' string mode.
At large $d$, there are actually many `tachyons'
and `massless modes' and hence
there are many options for the state of the spatial string
coordinates $Y_i$.

It is interesting to note that the two possible
signs of the $2d$ cosmological constant correspond
to the two inequivalent directions in which one can
push the bosonic string tachyon.  The effective
potential of the bosonic string has the structure
\be\label{Veff}
  \VV_{\rm eff} = -c_2 \TT^2 +c_3 \TT^3 +\ldots
\ee
(with $c_2$ and $c_3$ positive);
negative $2d$ cosmological constant $\mu>0$ 
corresponds to a tachyon field $\TT$ exponentially
growing in time in the direction of
what might be a metastable minimum at positive $\TT$,
while positive $2d$ cosmological constant $\mu<0$
has the tachyon growing in what naively looks
like an unbounded direction.
Of course, the higher order corrections to $\VV_{\rm eff}$
are order one, so one cannot say reliably that there is a local
minimum, or an unbounded direction. 

In the worldsheet theory, the positive tachyon condensate
acts as a barrier that prevents test strings from exploring
the region of large positive $\TT$ -- the classical solution
\pref{adssoln} suggests that strings try to annihilate one
another, since $\tau=\pm\infty$ both correspond to
early time $\phi\to-\infty$ in target space.
On the other hand, a negative tachyon condensate wants to
push test strings toward the asymptotic future in finite 
worldsheet conformal time.  Note that this type of
tachyon condensate is absent in the fermionic string
(\ie\ the type 0 theory), where the two signs of $\mu$
are actually equivalent -- changing the sign of $\mu$ is
undone by a chiral R-parity transformation on the worldsheet.
The physics there is always that of AdS gravity since only this sign
of cosmological term is compatible with worldsheet supersymmetry.

In the analogous open string problem,
one has similar issues, and the full theory {\it does} have
a local minimum of the tachyon effective potential -- 
the closed string vacuum.  In this regard, 
note also that the two signs of open string tachyon condensate
also have very different physics 
\cite{Sen:2002nu,Larsen:2002wc},
with an interpretation similar to the closed string
dynamics above.
Some aspects of the worldsheet physics 
of the $\mu<0$ boundary cosmological term in the bosonic string
were explored in \cite{Gutperle:2003xf}
(for a particular inhomogeneous condensate,
see \cite{Larsen:2002wc} and section \ref{LSGmodel} below).
One may be able to ascribe some of the pathologies
of their results to the fact that the open string
condensate is heading in the bottomless direction
of the spacetime effective potential.
It would be interesting to make a similar analysis
of the $\mu>0$ theory, especially since this is the
case relevant to the superstring.  
With this sign of boundary cosmological term,
the boundary potential `repels' the boundary from the future
$\phi\to\infty$, which is a way of seeing that open strings
are absent from the endpoint of the condensation process.
Of course, absent a bulk tachyon the dynamics in the interior
of the worldsheet is free to propagate to the future,
so the most likely outcome of the dynamics is
that the two ends of the string `annihilate' one
another and make a closed string, see figure \openstr.

\begin{figure}[ht]
\begin{center}
\[
\mbox{\begin{picture}(250,140)(0,0)
\includegraphics{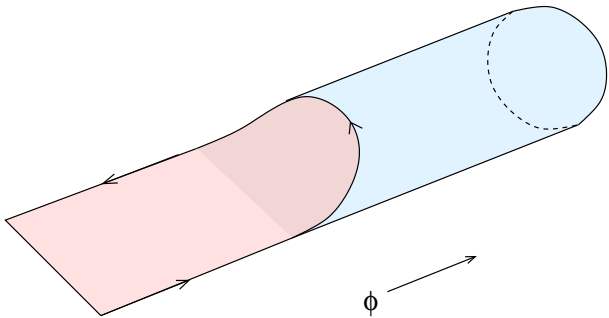}
\end{picture}}
\]
\caption{\it 
The boundary of a test open string propagating in an open string
tachyon background is `repelled from the future'.
The ends of the string find each other and annihilate,
leaving a free closed string whose propagation to the
future is unhindered.
}
\end{center}
\end{figure} 

In the case of open string tachyon condensation
(say in the superstring, where we don't have to worry
about issues of bulk closed string tachyons), there is
by now a fairly well-developed picture of the physics
of the condensate at late times from a variety
of points of view
\cite{Sen:1998sm,Sen:1999mg,Sen:1999nx,Harvey:2000na,Sen:2002nu}.
In a sense, open string boundaries are confined by the boundary potential,
as we have seen above, and at weak string coupling one expects
to find the closed string vacuum plus radiation. 
If the initial unstable brane was space-filling, 
one has a finite energy density of closed strings, 
leading to an FRW cosmology; if the initial unstable brane
is localized, its decay emits a spherical pulse of closed string 
radiation that travels outward to spatial infinity.

For closed string tachyons, the picture is less clear.
The endpoint of bulk closed string tachyon condensation
is unknown, although there have been speculations
\cite{Gutperle:2001mb,David:2001vm} that condensation
of the type 0 tachyon in the critical dimension $d=9$ is related
to type II string theory; if so, one would find a cosmological
solution of type II, not the type II vacuum, as the
condensate and the strings that are produced by it
source the Einstein equations.  In any event, there is
no known limit of such bulk tachyon condensation where
there is a controlled understanding of the late-time behavior.%
\footnote{Although the parallels with the open string case,
and the fact that perturbative strings are repelled
from the region of forming condensate,
suggest a phase where closed
fundamental strings are confined.}
Certainly the initial exponential growth of the tachyon field
does not carry on forever.  
Thus the Liouville action
above is not valid for all $\phi$, rather it breaks down
whenever strings explore the region of large $\phi$.

In the fermionic string, the worldsheet
cosmological constant is always negative,
thus giving rise to AdS cosmologies; as in the open string case,
the worldsheet boundary is repelled from the region
of large positive $\phi$.
There might be a self-consistent treatment
of some aspects of string propagation in the developing condensate,
because strings do not explore the region where we do not
trust the condensate.
However, the homogeneous tachyon condensate
describes AdS geometries and thus
does not give rise to inflation.  What we need is a nontrivial
matter potential, suitably chosen to give slow-roll inflation.


\section{\label{LSGmodel}Inhomogeneous tachyon condensates}

Tachyon condensates that are inhomogeneous in target space
describe gravitationally dressed matter potentials;
suitable choices of matter potential give rise to 
interesting two dimensional models of inflation.
Consider, for instance, the gravitational dressing
of a cosine matter potential -- the Liouville-Sine-Gordon model
\be\label{Slsg}
  \CS_{\sst\rm LSG} = 
	\frac{1}{4\pi}\int \! d\tau d\sigma\,\sqrt{-\hat g}\Bigl(
	\hf(\hat\nabla\phi)^2+\hf{Q}R(\hat g)\phi
	-\hf(\hat\nabla X)^2
	+\frac{\mu}{2\alpha^2}\,e^{\alpha\phi}\cos kX\Bigr)\ ,
\ee
where we have suppressed a further $d_\perp$ scalar fields $Y$
in which the tachyon condensate is homogeneous.
The conditions of conformal invariance are now
\bbb
  Q &=& \sqrt{(d_\perp-24)/3}
\nonumber\\
  \alpha &=& -\hf Q+\sqrt{\coeff14 Q^2+2-k^2}\ .
\label{lsgconf}
\eee
For sufficiently small $k$, the matter potential is very flat
and the conditions of slow-roll inflation are well satisfied.
Note also that $\alpha\sim\gamma\sim \sqrt{12/d_\perp}$
in the semiclassical limit of large $d_\perp$.

To compare the dynamics of this model with that of
inflation, it is useful to write the equations
of motion in synchronous gauge.  Let us define the `proper time'
\be\label{proptime}
  dt=e^{\alpha\phi/2}d\tau\equiv\aa\,d\tau\ ,
\ee
the local `Hubble parameter'%
\footnote{Note that this is not quite the expansion of the scale
factor $a=\exp[\gamma\phi/2]$; rather, one has $a=\aa^{\gamma/\alpha}$.
The use of $\aa$ simplifies the equations of motion.}
\be\label{pseudoH}
   H=\dot \aa/\aa = \hf \alpha\dot\phi
\ee
(where dot denotes $t$ derivative),
and the matter potential 
\be\label{mattpotl}
  V(X)=\frac{-\mu}{2\alpha^2}\;\cos(kX)\ .
\ee
Then the equations of motion and the Hamiltonian constraint 
become (for spatially homogeneous fields, and setting $\hat R=0$)
\bbb
  H^2 &=& \frac{\alpha^2}{2n}\,
	\Bigl(\hf {\dot X}^2+V(X)-\aa^{-2}(\coeff18 Q^2+1)+\rho_\perp\Bigr) 
\nonumber\\
  \dot H &=& -\frac{\alpha^2}{2}\;
	\Bigl(\hf {\dot X}^2 -\aa^{-2}(\coeff18 Q^2+1)
		+\rho_\perp+P_\perp\Bigr)
\label{friedmann}\\
  0 &=& \ddot X+n\,H\,\dot X+V'(X) 
\nonumber
\eee
with $n=1$ the number of spatial dimensions.
For $n$ spatial dimensions, these are precisely 
the Friedmann equations of 
a spatially homogeneous cosmology if we substitute
$\hf\alpha^2\to 8\pi G_{n+1}$, where $G_{n+1}$
is Newton's constant.%
\footnote{Note that the zero-point energy $-\frac18 Q^2-1$
plays the role of spatial curvature.
The sign is appropriate to a closed spatial universe in 
higher dimensions.}  

The conditions for slow-roll inflation
are then that the dimensionless parameters
\bbb
  \epsilon &\equiv& \half\, \frac{2}{\alpha^2}\Bigl(\frac{V'}{V}\Bigr)^2
\nonumber\\
  \eta &\equiv& \frac{2}{\alpha^2}\Bigl(\frac{V''}{V}\Bigr)
\label{slowroll}
\eee
have magnitude much smaller than one.  
For the cosine matter potential, this amounts to
\bbb
  \epsilon &=& (k/\alpha)^2 \tan^2 kX  \ll 1
\nonumber\\
  -\eta &=& 2(k/\alpha)^2 \ll 1\ ;
\label{LSGslowroll}
\eee
both are well satisfied for $k<<\alpha$ and an initial
matter distribution that starts near the top of the matter potential.%
\footnote{The field $X$ can be localized on scales larger than
the string scale (which we are setting equal to one).
For $k<<\alpha$ this will always be the case, especially
in the weak coupling regime of gravity $\alpha\sim\sqrt{3/d_\perp}$.}

For the special value $k=\alpha=\coeff14(-Q+\sqrt{Q^2+16})$, 
the classical theory can be solved exactly.%
\footnote{While only this value allows exact solution
to the classical equations of motion, we expect the
behavior for other (sufficiently small) values of $k$
to be similar.}
Unfortunately, the second of the slow roll conditions
\pref{LSGslowroll} is not satisfied
for this value of $k$; nevertheless, our hope is that
one may still extract useful lessons from the analysis of
this case.  While the condition $\epsilon\ll 1$ is 
the requirement that the cosmological expansion is dominated by
the inflaton potential rather than its kinetic energy,
the second condition $\eta\ll 1$ is to some extent
a convenience; it prescribes the circumstance
in which the inflaton motion is friction dominated,
so that $\ddot X\ll H\dot X$.%
\footnote{More precisely, $\eta$ not small does not 
preclude the inflationary phase; however, it does affect
observables such as the spectral tilt.  This level
of detail is beyond the scope of the present investigation.}
Precisely this situation, namely
$\epsilon\ll 1$ but $|\eta|\sim 1$, was studied
in \cite{Kallosh:2001gr,Linde:2001ae},
where it was termed `fast-roll' inflation.
Estimates there indicated that one can achieve a large
expansion of the scale factor without undue fine tuning;
indeed, explicit exact solutions below will support that analysis.

Writing $\Phi=\phi+iX$, the action separates for $k=\alpha$ into
a Liouville theory for $\Phi$ plus that of its complex conjugate:
\be\label{complexliou}
  \CS = 
	\frac{1}{4\pi}\int \! d\tau d\sigma\,\sqrt{\hat g}\Bigl(
	\coeff14(\hat\nabla\Phi)^2+\coeff14{Q}R(\hat g)\Phi
	+\frac{\mu}{4\alpha^2}\,e^{\alpha\Phi}\;+\;{\rm c.c.}\Bigr)\ .
\ee
The classical equations of motion thus have the general solution
\pref{gensoln}, where now we allow the functions $A(x^+)$ and $B(x^-)$
to be generically complex.  

For example, the zero energy solution
\be\label{complexflat}
  e^{\alpha\Phi} = \frac{4}{-\mu}\;\frac{1}{(\tau+ib)^2}
\ee
describes a scalar field $X$ that starts off at the top of its 
cosine potential at $\tau=-\infty$, rolls down
to the potential minimum at $\tau=0$, and then climbs back up to
the top of the potential as $\tau\to\infty$.
As this occurs, the geometry is nearly an expanding de~Sitter
space for large negative $\tau$; it reaches a maximum scale factor
$e^{\alpha\phi}=\coeff{4}{-\mu} b^{-2}$ at $\tau=0$, at which point it
enters a contracting phase and returns to zero scale factor
and asymptotically de~Sitter geometry as $\tau\to+\infty$.%
\footnote{Asymptotically de~Sitter in the sense of constant curvature;
one is not approaching conformal infinity of the global de~Sitter
geometry.}
Solving for the scale factor and matter field, we have
\bbb
  e^{\alpha\phi/2} &=& \aa(\tau) = \frac{1}{\sqrt{\tau^2+b^2}}
\nonumber \\
  \tan(\alpha X/2) &=& -b/\tau\ .
\label{compts}
\eee
If the lower bound on localizing the matter field $X$
is the string scale $\Delta X\sim 1$, then
the total amount of inflation is
\be
  \frac{\aa_{\rm final}}{\aa_{\rm initial}}\sim \frac{2}{\alpha b^2}\ .
\ee
The choice of $b$ amounts to the choice of the scale
factor $\aa$ at which one chooses to tune $X$ to lie
within a string length of the top of its potential.
By tuning $b$ to be small we can make this ratio
of scales as large as we like.

The analogues of the other two solutions exhibited in 
equation \pref{FRW} have similar interpretations.
The hyperbolic solution is
\bbb
  e^{\alpha\Phi} &=& \frac{4}{-\mu}\;
	\frac{(\varepsilon_\phi+i\varepsilon_{\sst X})^2}%
	{\sinh^2[(\varepsilon_\phi+i\varepsilon_{\sst X})\tau+ib]}
\nonumber\\
  e^{\alpha\phi} &=& \frac{8}{-\mu}\;
	\frac{(\varepsilon_\phi^2+\varepsilon_{\sst X}^2)}%
        {(\cosh[2\varepsilon_\phi \tau]
		-\cos[2\varepsilon_{\sst X}\tau+b])}\quad ;
\label{hyplsg}
\eee
again the geometry expands to a maximum scale, which
in the limit of small $\varepsilon_\phi$,
$\varepsilon_{\sst X}$, and $b$ reduces to 
\be\label{lsgmax}
  e^{\alpha\phi_{\rm max}} \sim 
	\frac{4}{-\mu}\;
	\frac{(\varepsilon_\phi^2+\varepsilon_{\sst X}^2)^2}%
	{ b^2 \varepsilon_\phi^2}\ .
\ee
Here both $\tau\to+\infty$ and $\tau\to-\infty$
give $2d$ Milne singularities.
The maximum scale factor achieved is finite unless we fine tune $b\to 0$,
so that the field reaches the top of the cosine potential
just as the scale factor starts inflating.
The BRST constraint
\be\label{lsgenergy}
  -\frac{\varepsilon_\phi^2}{2\alpha^2}
	+\frac{\varepsilon_{\sst X}^2}{2\alpha^2}
	+E_\perp-\frac{Q^2}{8}-1=0
\ee
relates the two zero mode energies.

Clearly there are a wide variety of tachyon profiles in
target space that one could investigate,
yielding a corresponding variety of inflaton potentials; 
one could imagine modelling any
of the popular inflationary scenarios
(\cf\ \cite{Lyth:1998xn} for a review), limited only
by one's ability to solve the dynamics.


\subsection{The fermionic string}

Suitable actions for the fermionic string 
are much the same as \pref{Sliou}, \pref{Slsg};
one simply substitutes the superspace integral $\int \!d^2xd^2\theta$
and derivative $\hat D$ for their bosonic counterparts,
and promotes $\phi$ and $X$ to worldsheet superfields.
Conformal invariance requires
\bbb
  Q &=& \sqrt{(d_\perp - 8)/2}
\nonumber\\
  \alpha &=& -\hf Q+\sqrt{\coeff14 Q^2+1-k^2}\ .
\label{superparams}
\eee
Elimination of auxiliary fields leads to a bosonic potential
\bbb
  \VV &=& G^{\phi\phi}(\partial_\phi\WW)^2
	+G^{\sst XX}(\partial_{\sst X}\WW)^2
\nonumber\\
	&=& \frac{\mu^2}{4\alpha^4}\,e^{2\alpha\phi}
		\left(-\alpha^2\cos^2 kX + k^2\sin^2 kX \right)
\label{bospotl}
\eee
where $G$ is the target space metric.
The minimum in $X$ yields an effective AdS cosmological
constant $-(\mu/4\alpha)^2$ (as mentioned above
in section \ref{STinterp}, this is required by supersymmetry), 
while the maximum corresponds
to a dS cosmological constant $+(\mu k/4\alpha^2)^2$.
Again the choice $k=\alpha=\coeff14(-Q+\sqrt{Q^2+8})$
leads to a classically integrable theory --
the complexified super-Liouville model.
Its classical solutions differ little from those discussed above.


\subsection{Winding modes and domain wall inflation}

Another scenario for inflation uses topological
defects as seeds for `eternal' inflation
\cite{Vilenkin:1994pv,Linde:1994wt}.
Since a scalar field at the core of such a defect
is pinned at the maximum of its potential,
if the characteristic size of the defect exceeds the
Hubble scale, the interior of the defect will inflate.
The Sine-Gordon-Liouville theory provides a simple
model of this sort as well -- we simply compactify
the scalar $X$ on a circle of radius $2/k$
(in string units) and consider winding strings.
A prototypical solution of this sort in the $\alpha=k$ model has
\be\label{windsoln}
  A(x^+)=\exp[(\varepsilon -iw)x^+]
	\quad,\qquad
  B(x^-)=\exp[(-\varepsilon -iw)x^-]\ ,
\ee
leading to
\be\label{windliou}
  \exp[\alpha\Phi] = 
	\frac{4}{-\mu}
	\frac{\varepsilon^2+w^2}{\sinh^2(\varepsilon \tau-iw\sigma)}\ .
\ee
At large $\tau$, one sees that the imaginary part $X={\rm Im}\,\Phi$
winds $w$ times.
For the component Liouville and matter fields 
$\phi= {\rm Re}\,\Phi$ and $X$, one finds
\bbb
  e^{\alpha\phi} &=& \frac{8}{-\mu}\;
	\frac{\varepsilon^2+w^2}{\cosh(2\varepsilon \tau)-\cos(2w\sigma)}
\nonumber\\
  \sin(\alpha X) &=& \frac{\sinh(2\varepsilon \tau)\sin(2w\sigma)}%
			{\cosh(2\varepsilon \tau)-\cos(2w\sigma)}
\label{reimpart}
\eee

\begin{figure}[ht]
\begin{center}
\[
\mbox{\begin{picture}(200,200)(10,0)
\includegraphics{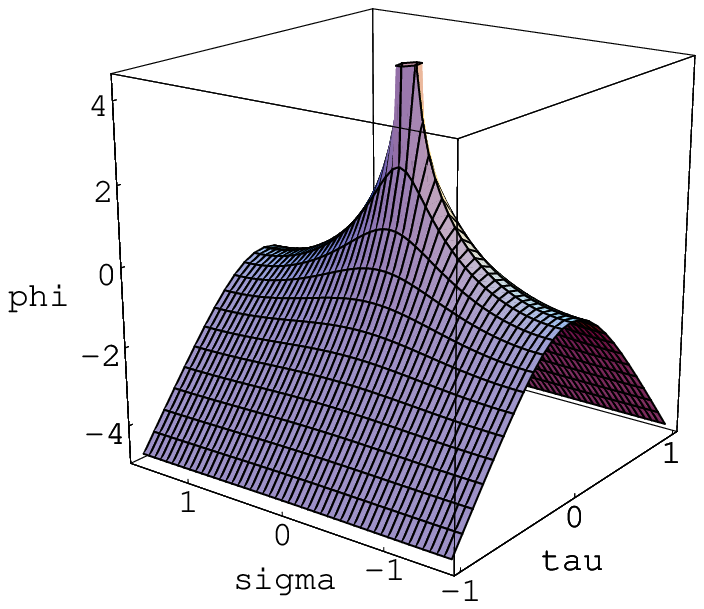}
\end{picture}
\begin{picture}(200,200)(-10,0)
\includegraphics{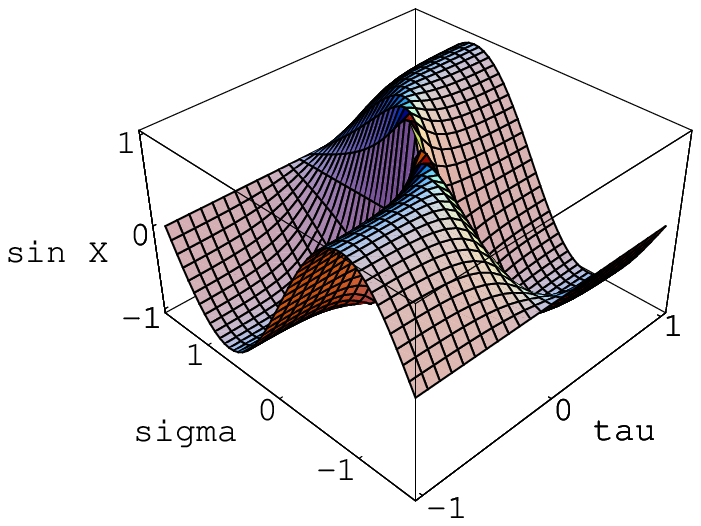}
\end{picture}
}
\]
\caption{\it
Plots of the scale factor $\phi$ and matter field $X$
during topological inflation.  At early and late worldsheet time $\tau$,
the matter field $X$ has one unit of winding.  The spacetime
picture, however, consists of a pair of oppositely
wound strings annihilating and leaving behind an unwound
string perched at a maximum of the cosine matter potential.
}
\end{center}
\end{figure}

From the plots of $\phi$ and $X$ in figure \topinf,
one sees that this solution describes a pair of oppositely
wound strings which annihilate, leaving an unwound string
perched at the top of the cosine potential.  Of course,
in the quantum theory one does not expect this late-time
(large $\phi$) behavior to be stable; rather, one
expects that the field $X$ falls off the hill in either direction.
It is interesting to note that, even though we naively
start off with a coordinate domain $\IR\times S^1$
that we wish to interpret as a single closed universe
cosmology, the actual dynamics grows a second `child universe'
that absorbs the domain wall core and allows it to recollapse
after fluctuations drive $X$ off its potential maximum.%
\footnote{Modulo issues of `eternal' inflation, which
we defer to section \ref{eternalchaos}.}
It would be interesting to explore whether such a mechanism
disallows eternal topological inflation in higher dimensions;
even so, such topological structures might be a natural way of
providing the initial conditions for inflation,
since the core of the defect is placed on the
top of the inflaton potential at the point in
time when the topological charge is shed.


\section{\label{localtach}Localized tachyon condensates}

While the above $2d$ models of inflation are simple and
appealing, there are several concerns that need to be addressed
arising from the target space interpretation in terms
of closed string tachyon condensation.
The unlimited exponential growth in time of the tachyon condensate
is unrealistic; at some point, the tachyon condensate grows
large enough that one cannot ignore its back-reaction
on the target space geometry.  Ultimately, one will need to know
the late-time state that the condensate evolves toward
and whether it even has a perturbative string interpretation.
The fate of bulk closed string tachyon condensates in
both bosonic and fermionic string theory is an open question.
In addition, unless one works in the critical dimension,
one should worry about the consistency of
propagation of a single string (our $2d$ cosmos)
in a background where the string coupling increases
linearly in target space time as we go to the past;
the `in' region is not under control.

An alternative to this admittedly murky situation
is provided by {\it localized} closed string tachyon condensates,
in the critical dimension.
The above difficulties are associated to the instability
of the target spacetime, which occur everywhere in space
with little understanding of the outcome.  However,
the test string which is our two-dimensional cosmology
can be localized in the target space in a region of localized
instability; away from this region,
string theory is stable.  
In a sense, one can view the two-dimensional cosmology
as a kind of scattering of a test string off of
the decaying localized defect.%
\footnote{L. Susskind \cite{Susskind:2003kw}
has expressed a similar point of view on de~Sitter
cosmology in a related context.}
String theory in the supercritical dimension 
always has bulk tachyons, even with a chiral GSO 
projection, \cf\ \cite{Myers:1987fv}.
Therefore, we must work in the critical dimension $d+1=10$
of the type II string, in a background with a localized
closed string tachyon.  Such backgrounds have been explored 
in various contexts in 
\cite{Adams:2001sv,Vafa:2001ra,Harvey:2001wm,David:2001vm},
as well as many further works.  
In these examples, 
the target space contains an unstable defect, and the closed string
tachyon is a description of its initial stages of decay.
Of course, the critical 
dimension is the limit $Q\to 0$, $\gamma\to 1$, which
is a `strong coupling limit' from the point of view of
the worldsheet gravity theory.  What this means in practice
is that fluctuations of the timelike Liouville field $\phi$
are of the same order as those of the matter fields,
of order one at the string scale.  We will also need
to take care that slow-roll conditions can be satisfied.

As argued first in \cite{Adams:2001sv},
localized closed string tachyon condensation 
leads to a stable remnant, or flat spacetime,
with an outgoing pulse of radiation.
As viewed by a test string that remains near the origin,
this process would appear as an exponential growth 
of the timelike Liouville field coupled to a localized matter
perturbation, which then saturates and relaxes to a smooth
final configuration.  In other words, the $2d$ cosmological
constant dynamically relaxes to zero at large scale factor $\phi$!
We thus find an explicit and self-consistent realization 
of a `wormhole' style mechanism \cite{Coleman:1988tj} for relaxing the
cosmological constant.%
\footnote{The idea that the $2d$ cosmological constant
is relaxed by `wormholes' or `baby universes' was explored already
in the work of \cite{Cooper:1991vg,Cooper:1991zc}. 
At that time, no definite conclusions
could be drawn due to the state of our understanding 
of bulk closed string tachyon condensation.  Our addition
to the discussion is to consider localized tachyons in an
otherwise stable closed string background.}

Consider, for example, the throat models introduced in section 5 of
\cite{Harvey:2001wm}.  The CFT background is 
\be\label{throatcft}
  \IR^{5,1}\times\Bigl(\frac{SL(2)_{kn}}{U(1)}\times
	\frac{SU(2)_{kn}}{U(1)}\Bigr)/\IZ_k\ .
\ee
The chiral GSO projection demands $n\in 2\IZ+1$.
The $SU(2)/U(1)$ factor in the background \pref{throatcft}
is quantum equivalent to an $\NN=2$ supersymmetric 
Landau-Ginzburg model for a chiral superfield $X$%
\cite{Martinec:1989zu,Vafa:1989uu},
with a superpotential
\be\label{LGpotl}
  \WW_{\sst X}=X^{kn}
\ee
The cigar CFT $SL(2)/U(1)$ is quantum equivalent
\cite{Giveon:1999px}
to a {\it spacelike} $\NN=2$ supersymmetric
Liouville theory, not to be confused
with the {\it timelike} $\NN=1$ Liouville theory we will use
to describe the time evolution of the tachyon condensate.
We describe this spacelike Liouville theory using
a chiral $\NN=2$ superfield $Y$ with the superpotential
\be\label{ntwoliou}
  \WW_{\sst Y}=\tilde\mu\, e^{-Y/\tilde Q}
\ee
where $\tilde Q^2=2/kn$ is the linear dilaton slope in the spatial Liouville
direction $\varphi={\rm Re}\;Y$, and $\tilde\mu$ determines the string
coupling at the tip of the cigar. 

This theory has a moduli space, related to the moduli space
of the throat of $k$ fivebranes that remains after tachyon 
condensation \cite{Harvey:2001wm}.
This moduli space is explored by deforming the
$\NN=2$ superpotential \pref{LGpotl} plus \pref{ntwoliou} to
\be\label{moddefW}
  \WW=X^{nk}+\tilde\mu\, e^{-Y/\tilde Q}+
	\sum_{j=2}^{k-1}\lambda_j X^{n(k-j)}e^{-{\tilde Q\over2}jn Y}
\ee
Note that all terms in \pref{moddefW} are invariant under the
$\IZ_k$ symmetry in \pref{throatcft}, which acts as
\be\label{znsym}
  X\to e^{2\pi i\over k}X\ ;\;\;
	Y\to Y-2\pi i\tilde Q n\ ;\;\;
	\theta\to e^{2\pi in}\theta\ ;\;\;
	\bar\theta \to\bar\theta
\ee
The parameters $\lambda_j$ in \pref{moddefW} determine the locations
of the underlying fivebranes; for $\lambda_j=0$,
they are located at the origin in $X$ and arranged in
a $\IZ_k$ symmetric fashion on the ${\rm Im}\,Y$ circle.

The modes corresponding
to these deformations have normalizable wavefunctions
concentrated down the throat (at large negative ${\rm Re}\,Y$).  
When the $\lambda_j$ are turned on, 
they are localized away from the origin in $X$:
The $\NN=2$ superpotential generically
has $k$ distinct $n$-fold degenerate extrema
where $\NN=2$ supersymmetric ground states are localized.
There are also tachyonic deformations surviving
the $\IZ_k$ projection \pref{znsym}.  These are 
perturbations $\delta\WW= \mu_\ell X^{k\ell}$;
again only $\ell\in 2\IZ+1$ survives the GSO projection.
In order to satisfy the BRST constraints, we dress these perturbations
with time dependence to put them on-shell.
The time dependence breaks the global worldsheet $\NN=2$
supersymmetry; we can only write the tachyon background 
in terms of an $\NN=1$ worldsheet superpotential
\be\label{throattach}
  \delta(\WW+\bar\WW) = 
	\sum_{\ell=1}^{n-1}(\mu_\ell X^{k\ell} +{\rm c.c.})
		\, e^{-\alpha_\ell \phi}\ .
\ee
A nontrivial potential is generated for the test string
due to the competition between the moduli-induced potential \pref{moddefW},
which wants to localize the string near one of the fivebrane sources,
and the tachyon-induced potential \pref{throattach},
which wants to localize the string elsewhere (\eg\ the origin
if we only turn on $\mu_1$).  By suitable tuning of $\tilde\mu$,
the $\mu_\ell$, and the $\lambda_j$, one can arrange that
the test string is initially trapped in a metastable minimum 
of the potential at large $|X|$ for an extended period of time,
generating inflation.
Eventually, the tachyon potential \pref{throattach} grows at large $\phi$ 
and dominates the dynamics -- the scalar $X$ rolls down
from this large value to a minimum at $X=0$.
The conditions of slow roll will be satisfied for
sufficiently large initial values of $X$
(large relative to the string scale), for which 
the `inflaton' rolls down from large values and
starts off with a wavefunction that is localized on
the string scale.
After this period of inflation, at late time the test string
settles down with some value of the worldsheet energy $E_\phi$
corresponding to an expanding FRW-like cosmology at late times,
since the growth of the tachyon condensate saturates
and blows out to spatial infinity (far from $X\sim 0$)
leaving a vanishing potential for $\phi$, when $\phi$ is large
and $X$ is near zero.%
\footnote{It is possible that the condensation process
excites a motion of the fivebranes on their moduli space
that causes them to collide; coincident fivebranes generate
large string coupling and the perturbative approximation
breaks down.  It will be interesting to see if
this happens; the analysis is left to future work.
Regardless of the result, we can arrange for the fivebranes
to be initially well-separated, so that
the collision occurs a long time after the exponential
growth of the tachyon condensate,
and therefore the period of worldsheet inflation, shuts off.}

We believe that one can also form interesting inflationary
models using the nonsupersymmetric orbifold constructions of
\cite{Adams:2001sv,Vafa:2001ra,Harvey:2001wm}, although
we have not investigated them in much detail.  The key feature
allowing for an inflationary epoch in both these models 
and the fivebrane throat models discussed above,
is the the ability to describe string scale geometry
through the use of an effective scalar {\it potential}
rather than a nonlinear sigma model.  Prime examples of this
are the equivalence of $SL(2)/U(1)$ to $\NN=2$ Liouville,
and of $SU(2)/U(1)$ to a Landau-Ginsburg model --
geometry is entirely converted into a scalar potential.
It would be interesting if this feature extended to higher 
dimensions, so that the effective fields that govern the
theory of the early universe are best described
in terms of dynamics in an effective potential, which depends
on the epoch in spacetime of interest and may be absent at late times.


\section{\label{Qeff}Quantum effects}

In this section, we explore three commonly considered
semiclassical phenomena: The quantum stress tensor
of matter in a time-dependent gravitational background;
the minisuperspace approximation to the wavefunction
of the universe; and the gravitational thermodynamics
of de~Sitter space.  In addition, we discuss string
pair creation in closed string tachyon backgrounds.

\subsection{The stress tensor and particle production}

A standard method of assessing the effects of quantum
fluctuations on inflation is to compute the semiclassical
fluctuations around the classical slow-roll inflation
solution.  In the complexified Liouville model of section
\ref{LSGmodel}, we are in a considerably better situation --
it may be possible to analyze the full quantum dynamics,
given the fact that the theory is classically integrable.
We leave this exercise to future work.
As a first step, let us perform the standard semiclassical
analysis on this model 
(\cf\ \cite{Birrell:1982ix} and references therein).
This approach should be valid in the limit of large transverse
dimension $d_\perp$, where the fluctuations of the geometry
are parametrically suppressed by $1/d_\perp$.

For a conformal field theory of central charge $c$,
the conformal anomaly relates the expectation value 
$\vev{T_{\alpha\beta}}$ of the stress tensor 
$T_{\alpha\beta}=2\pi\frac{\delta \CS}{\delta g^{\alpha\beta}}$
in the metric $e^{\gamma\phi}\hat g$
to its expectation value $\vev{\hat T_{\alpha\beta}}$ in the
metric $\hat g$:
\bbb
  \vev{T_{+-}} &=& \vev{\hat T_{+-}}
 - \frac{c}{24} g_{+-}\,R
\nonumber\\
  \vev{T_{++}} &=& \vev{\hat T_{++}}
 - \frac{c\gamma^2}{12} \Bigl(\hf(\partial_+\phi)^2
  -\hf{Q}\partial_+^2\phi\Bigr)
\label{anomaly}\\
  \vev{T_{--}} &=& \vev{\hat T_{--}}
 - \frac{c\gamma^2}{12} \Bigl(\hf(\partial_-\phi)^2
  -\hf{Q}\partial_-^2\phi\Bigr)\ .
\nonumber
\eee
Thus, begin with the theory on $2d$ Minkowski spacetime
and conformally transform to the (flat) cylinder; this gives
a stress tensor on the cylinder
\be\label{cyltens}
   \hat T_{+-}=0\quad ,\qquad
   \hat T_{++} = \hat T_{--} = -\frac{c}{24}\ ;
\ee
then perform the Weyl transformation to the metric \pref{twodmetric}.
For instance, the global de~Sitter metric \pref{globalds}
gives 
\be\label{globaldstress}
   T_{+-}=-\frac{c}{24}\;\frac{\epsilon^2}%
	{\sin^2 (\epsilon \tau)}\quad ,\qquad
   T_{++} = T_{--} = -\frac{c}{24}\;(1-\epsilon^2)\ .
\ee
In the semiclassical approximation, this agrees
with the analysis of section \ref{liourev},
since $\frac1{24} c = \frac18 Q^2$.
Similarly, the stress tensor for the geometry
\pref{milneds} (similarly \pref{flatds}) is obtained
by the analytic continuation 
$\varepsilon=i\epsilon$ (similarly $\epsilon\to0$)
\be\label{pastcasimir}
  T_{+-}=-\frac{c}{24}\;\frac{\varepsilon^2}%
	{\sinh^2 (\varepsilon \tau)}\quad ,\qquad
  T_{++} = T_{--} = -\frac{c}{24}\;(1+\varepsilon^2)\ .
\ee
The shrinking proper size of the spatial circle 
in the far past leads to a diverging Casimir energy
for conformal matter, as for instance the scalar quantity
$T^{\alpha\beta}T_{\alpha\beta}$ diverges as $\tau\to-\infty$.
Thus, even though the total stress tensor of matter plus ghosts
plus Liouville gravity vanishes, individual ingredients
have diverging stress-energy in this sense.
Of course, if we do not interpret the Liouville field as the
scale factor of the two-dimensional metric, then we are free to use
the standard CFT vacuum on the flat cylinder \pref{cyltens};
it is only the attempt to interpret the target space time
as the $2d$ metric that causes difficulty;
even then, it is the choice of observable
such as $T^{\alpha\beta}T_{\alpha\beta}$ that is problematic,
and not necessarily the CFT state that we have defined.

The complexified Liouville theory of section \ref{LSGmodel}
leads to a more intricate geometry, and the corresponding
quantum stress tensor is now time-dependent.  For example, the
zero energy solution \pref{complexflat}
leads to a stress tensor
\be
\label{flatstress}
  T_{+-} = -\frac{c}{24}\; \frac{\tau^2-b^2}{(\tau^2+b^2)^2}
\quad,\qquad
  T_{++}=T_{--} = -\,\frac{c}{24}\;
	\Bigl(1+\frac{b^2}{(\tau^2+b^2)^2}\Bigr)\ .
\ee
The source of this time-dependence is the dynamical
evolution of the inflaton $X$, which trades energy
with the gravitational field through the gravitational
dressing of the cosine potential.

In the positive energy case, we specialize 
for simplicity to $\varepsilon_{\sst X}=0$.
The Liouville stress tensor is
\bbb
  T_{+-} &=& -\frac{c}{24}\;\frac{\varepsilon_\phi^2[
	\cos(2b)\cosh (2\varepsilon_\phi\tau)-1]}%
               { [\cosh (2\varepsilon_{\phi}\tau) - \cos(2b)]^2}
\label{cosmostress}\\
  T_{++} &=& T_{--}=-\frac{c}{24}\Bigl(1+
	\frac{\varepsilon_\phi^2\,[\sinh^2(2\varepsilon_\phi\tau)%
	-2\cos(2b)\cosh(2\varepsilon_\phi\tau) -1]}%
		{[\cosh(2\varepsilon_\phi\tau) - \cos(2b)]^2}\Bigr)
\nonumber
\eee
which has the same qualitative features as the zero energy solution.

Linearized analysis of the analogous four-dimensional
problem also leads to an apparent negative stress-energy
\cite{Tsamis:1996qq,Abramo:1997hu} along the lines of  
\pref{pastcasimir}, \pref{flatstress}.
These works interpreted this result 
as a possible back-reaction that `slows' inflation,
or `screens' the cosmological constant.
The Liouville-Sine-Gordon model throws some doubt
on this interpretation.  
The worldsheet stress-energy `induced' by the Weyl anomaly 
\pref{anomaly} 
is in fact just the expectation value of the quantum
Virasoro (or BRST) constraints on the state of the string.
The quantum matter stress-energy simply determines the
class of de Sitter solution that the cosmology follows.
There is no dynamical `screening'
of the tachyon condensate at late times; 
the relations \pref{anomaly} are
related to the characterization of the string state,
and do not alter couplings in the worldsheet action
such as the value of the cosmological constant.

Another important aspect of quantum field theory in curved spacetime
concerns the fact that, generically, a time-dependent metric
will cause the production of field quanta
(\cf\ \cite{Birrell:1982ix} and references therein).
The generation of such fluctuations during inflation
and their subsequent back-reaction on the geometry
provides the seeds of large-scale structure,
one of the landmark successes of inflationary theory
(for reviews, see \eg\ \cite{Lyth:1998xn,Liddle:2000cg}).
The present $2d$ models provide a laboratory to study
concretely and quantitatively the mechanisms of generation
and back-reaction of quantum fluctuations, 
as well as related issues in quantum cosmology.

The great advantage of two-dimensional models is the simple nature
of $2d$ gravity.  However, the Liouville-Sine-Gordon model
is in this regard perhaps too simple~-- 
apart from the inflaton field,
all other matter is massless and conformally coupled, 
and therefore only sensitive
to the geometry through the Weyl anomaly.  In particular,
there will be no particle creation -- all transverse fields
remain happily in their conformal vacuum \cite{Birrell:1982ix}
(or whatever initial state we decide to put them in). 
There is nobody present to disturb the state
and observe the generation of fluctuations!
In order to recover the standard inflationary paradigm,
one needs either to introduce massive matter (appropriately
gravitationally dressed), or some class of observables such as a
collection of comoving particle detectors.
These will then register the presence
of fluctuations in the inflaton field,
which decohere into classical fluctuations of the geometry,
and lead to the standard picture of structure formation.
This lack of decoherence, together with the exact integrability
of the model, may explain why the solutions 
\pref{complexflat},\pref{hyplsg} of the complexified
Liouville theory are time-symmetric
and recollapse in precisely the same way that they expand.
One would expect that anything that disturbs the state,
such as interaction of the inflaton with massive matter,
or the insertion by hand of some observable to measure the 
state, will destroy the time symmetry and lead
to a recollapsing state rather different
in character from the quiescent initial state.%
\footnote{See related comments in
\cite{Hollands:2002yb,Hollands:2002xi}.}


\subsection{The wavefunction of the universe}

Another common approximation scheme is the so-called
minisuperspace approximation, which suppresses the
non-zero mode fluctuations of the fields.  While perhaps
less well-justified than the semiclassical approximation
above, it often yields fruitful insights into the dynamics.
In particular, it captures many of the qualitative features
of Liouville theory in subcritical dimensions $d<25$,
\cf\ \cite{Ginsparg:1993is}.  We begin with an examination
of the Liouville model in the supercritical dimension
$d>25$, and then turn to the complexified Liouville 
model of inflaton dynamics.

The Schr\"odinger (Wheeler-DeWitt) equation 
of Liouville quantum mechanics is
\be\label{wdweq}
  \hf(L_0+\bar L_0)\Psi_{\omega}(\phi) 
	= \Bigl[\half\Bigl(\frac{\partial}{\partial\phi}\Bigr)^2
		-\frac{\mu}{8\gamma^2}\,e^{\gamma\phi}
		-\frac{Q^2}{8}\Bigr]\Psi_{\omega}(\phi)
	= (1-E_{\sst\rm matter})\Psi_{\omega}(\phi)\ .
\ee
The solutions to this equation are Bessel functions,
whose asymptotics are
\bbb
  J_\nu (z)\sim \cases{\coeff1{\Gamma(\nu+1)}(\hf z)^\nu &\qquad $(z\to0)$\cr
		& \cr
  \sqrt{\coeff{2}{\pi z}}
        \cos(z-\coeff{\pi}{2}\nu-\coeff{\pi}{4})
                &\qquad $(z\to\infty)$}
\label{Jbessel} \\
\nonumber\\
  N_\nu (z)\sim \cases{
	\coeff{- \Gamma(\nu)}{\pi}
	(\hf z)^{-\nu} &\qquad $(z\to0)$ \cr
		& \cr
	\sqrt{\coeff{2}{\pi z}}
	\sin(z-\coeff{\pi}{2}\nu-\coeff{\pi}{4})
		&\qquad $(z\to\infty)$}
\label{Nbessel} 
\eee
The de~Sitter solutions which are pure positive frequency 
plane waves in the far past, 
\be\label{smallscale} 
  \Psi_{\omega}^{\it in}(\phi)\sim 
	\frac{1}{\sqrt{2\omega}} e^{-i\omega\gamma\phi}\ ,
\ee
correspond to the classical trajectories \pref{milneds}.
They have the minisuperspace wavefunction
(see \eg\ \cite{Strominger:2002pc,Gutperle:2003xf})
\bbb
  \Psi_{\omega}^{\it in}(\phi) &=& 
	\frac1{\sqrt{2\omega}}\,
	\Bigl(\frac{-\mu}{4\gamma^4}\Bigr)^{i\omega}
	\Gamma(1-2i\omega)\,
	J_{-2i\omega}({\sst \sqrt{\frac{-\mu}{\gamma^4}}}\,e^{\gamma\phi/2})
\nonumber\\
	\omega^2 &=& \frac{2}{\gamma^2}(E_{\sst\rm matter}-1-\coeff18 Q^2)
		=\Bigl(\frac{\varepsilon}{\gamma^2}\Bigr)^2\ .
\label{wdwsoln}
\eee
The classical trajectory is of course 
the WKB trajectory, which is a good approximation
for sufficiently large $\omega$.
For example, at small scale factor the solution
\pref{wdwsoln} behaves as \pref{smallscale},
a plane wave of momentum $\omega\gamma$,
and the corresponding classical solution \pref{milneds} 
of the same Liouville energy $E_L$ has the
same momentum, $\varepsilon=\omega\gamma^2$.
Similarly, the solutions \pref{globalds}
describing classical pair production are given by
the analytic continuation of $\omega$ to imaginary values,
so that the wavefunction vanishes exponentially in $\phi$
at small scale factor, and has exponentially large momentum
$\Pi_\phi\sim \pm\exp[\gamma\phi/2]$
at large scale factor.%
\footnote{Both signs appear because a given scale factor
appears twice in the classical solution -- once during the
contracting phase, and again in the expanding phase.}

Although the wavefunction is analytic in $\mu$,
the analytic continuation to AdS cosmological constant $\mu>0$ 
does not naively give the wavefunction corresponding to
the AdS solution \pref{adssoln}.  The analytic continuation
sends $J_\nu(z)\to J_\nu(iz)= iI_\nu(z)$, which blows up
exponentially at large 
$z={\sst \sqrt{\frac{-\mu}{\gamma^4}}}\,e^{\gamma\phi/2}$, 
and has only an expanding 
and not a contracting phase at small $\phi$ 
-- certainly not what
we expect on the basis of the classical solution \pref{adssoln},
which expands to a maximum value and then recontracts.
Rather, the AdS solution \pref{adssoln} corresponds to the
Bessel function 
\be
\label{Kbessel} 
  K_\nu(z) = \coeff{i\pi}{2}e^{i\nu\pi/2}[J_\nu(iz)+iN_\nu(iz)]
	\sim \cases{\frac{\pi}{2\sin\pi\nu}\left(
		\frac{(z/2)^{-\nu}}{\Gamma(1-\nu)}
		-\frac{(z/2)^\nu}{\Gamma(1+\nu)}\right)
		&\quad $(z\to 0)$ \cr
	& \cr
	\sqrt{\frac{\pi}{2z}}\; e^{-z} & \quad $(z\to\infty)$ \quad.}
\ee
This solution has both positive and negative frequency components 
at small scales, matching the fact that a given scale
factor reached classically appears both during expansion
out of the big bang
(when $\phi$ has positive momentum) and contraction
toward the big crunch
(when $\phi$ has negative momentum);
and it also vanishes exponentially for large scales,
corresponding to the fact that, for a given 
$\varepsilon=\omega\gamma^2$,
there is a maximum scale that is obtained before 
the universe recollapses.

The analytic continuation in $\mu$ of the wavefunction \pref{wdwsoln}
appears instead to be related to the static solutions 
\pref{globalads}, \pref{poincare}, and \pref{rindlerads}.
Each of these admits a formal periodic identification along
their timelike Killing vector to yield a spacelike cylindrical
worldsheet, and is the solution directly related by analytic
continuation of \pref{globalds}, \pref{flatds}, and \pref{milneds}
under $\mu\to -\mu$, together with $\tau\leftrightarrow\sigma$.%
\footnote{The worldsheet coordinates are of course invisible
to the Wheeler-DeWitt formalism.}
These classical AdS solutions have support at large scale factor
$\phi\to+\infty$, but now it is at {\it spatial} infinity
of the classical solution, and the metric is identified
along a {\it timelike} isometry.
This aspect of the wavefunctions -- having parts of the
wavefunction associated to closed universe cosmologies,
and parts of the wavefunction associated to spatially
noncompact regions with timelike identification --
is in some respects reminiscent of 
the Nappi-Witten geometries studied in 
\cite{Elitzur:2002rt}.

To summarize, there are two linearly independent solutions
to the Wheeler-DeWitt equation, and each can be
given an interpretation in terms of classical 
solutions in the WKB approximation.
The boundary conditions to be imposed on the
solution amount to the choice of a preferred linear combination
of these solutions, and it is not clear that 
Feynman boundary conditions (which are most
natural to the string theoretic interpretation)
are the most natural choice.
The imposition of Feynman boundary
conditions leads to exponentially growing wavefunctions
in the far future.
We have interpreted this behavior as being related to
static AdS geometries, but
clearly this feature needs to be better understood.

The complexified Liouville theory of section \ref{LSGmodel}
gives a trivial modification of \pref{wdweq}.
Take as the relevant Schr\"odinger equation for the
holomorphic coordinate $\Phi$
\be\label{Phiwdw}
  \hf(L_0+\bar L_0)\,\Psi_\omega(\Phi)
  	= \Bigl[\frac14\Bigl(\frac{\partial}{\partial\Phi}\Bigr)^2
		-\frac{\mu}{16\alpha^2}\,e^{\alpha\Phi}+{\rm c.c.}
		\Bigr]\Psi_{\omega}(\Phi)
	= -\frac{\alpha^2\omega^2}{4}\,\Psi_{\omega}(\Phi)
\ee
for complex $\omega$. 
For example, the tensor product of the wavefunctions \pref{wdwsoln}
\be\label{combsoln}
  \Psi_{\omega}^{\it in}(\Phi)
	{\Psi}_{\bar\omega}^{\it in}({\bar\Phi})
\ee
then satisfies the Schr\"odinger equation
\be
  \hf(L_0+\bar L_0)\Psi_{\omega}(\Phi)\Psi_{\bar\omega}(\bar\Phi) 
	= -\frac{\alpha^2}{4}(\omega^2+\bar\omega^2)
	\Psi_{\omega}(\Phi)\Psi_{\bar\omega}(\bar\Phi) \ .
\label{prodeq} 
\ee
In terms of the classical solutions of section \ref{LSGmodel},
one has $\varepsilon_\phi+i\varepsilon_{\sst X}=\omega\alpha^2$.

An important issue surrounds the choice of boundary conditions
on the wavefunction.  The particular choice \pref{combsoln}
does not satisfy the condition of periodicity 
under $X\to X+2\pi/\alpha$.  
Here it proves useful to employ Hankel functions 
\bbb
  H_\nu^{\sst(1)}(z) &=& J_\nu(z) + iN_\nu(z) 
\nonumber\\
  H_\nu^{\sst(2)}(z) &=& J_\nu(z) - iN_\nu(z) \ ,
\label{Hbessel}
\eee
whose properties under analytic continuation are
(\cf\ \cite{gradshteyn})
\bbb
  H_\nu^{\sst(1)}(e^{m\pi i}z) &=&
	\coeff{\sin(1-m)\nu\pi}{\sin\,\nu\pi}\; H_\nu^{\sst(1)}(z)
	-e^{-\nu\pi i}\;\coeff{\sin\,m\nu\pi}{\sin\,\nu\pi}\;
		H_\nu^{\sst(2)}(z) 
\nonumber\\
  H_\nu^{\sst(2)}(e^{m\pi i}z) &=&
	\coeff{\sin(1+m)\nu\pi}{\sin\,\nu\pi}\; H_\nu^{\sst(2)}(z)
	+e^{+\nu\pi i}\;\coeff{\sin\,m\nu\pi}{\sin\,\nu\pi}\;
		H_\nu^{\sst(1)}(z)
\label{hankeltransf}
\eee
for any integer $m$.
For simplicity, we will restrict our considerations to 
${\rm Im}\,\omega = \varepsilon_{\sst X}/\alpha^2=0$;
this is in any case the situation of most interest, where the
zero mode of the matter field is initially stationary.
Using these transformation properties, one may show that
\be\label{nicefn}
  \Psi_\nu(z,\bar z) = C_\nu\Bigl(
	e^{\nu\pi i}\;H_\nu^{\sst(1)}(z)H_\nu^{\sst(1)}(\bar z)
	-e^{-\nu\pi i}\;H_\nu^{\sst(2)}(z)H_\nu^{\sst(2)}(\bar z)
		\Bigr)
\ee
(where $C_\nu$ is a normalization)
is invariant under $z\to e^{m\pi i}z$. 
Appropriate wavefunctions for the complexified Liouville theory
are then $\Psi_\nu(e^{\alpha\Phi/2},e^{\alpha\bar\Phi/2})$,
where we have taken the liberty to absorb the factor
$\sqrt{-\mu/\alpha^4}$ into a shift of $\Phi$.
The asymptotics of $\Psi_\nu$ are
\be\label{psiasymp}
  C_\nu^{-1}\Psi_\nu(z,\bar z) \sim \cases{
	-\frac{i}{2\,\sin\,\pi\nu}\left[
		\frac{(z\bar z)^\nu}{(\Gamma(1+\nu))^2}
		+\frac{(z\bar z)^{-\nu}}{(\Gamma(1-\nu))^2}\right]
	&\qquad$|z|\to 0$ \cr
		& \cr
	-\frac{4i}{\pi\,|z|}\; \cos(z+\bar z) & \qquad $|z|\to\infty$\ .}
\ee
For $\nu=i\omega$, $z=\exp[\alpha\Phi/2]$,
we see that there is a unit reflection amplitude,
as in \pref{Kbessel} for 
geometries expanding from small scale factor.
However, there is also the behavior \pref{Jbessel}
appropriate to a global de~Sitter geometry
at large scale factor.
What is going on?  The scalar field $X$ in the
geometry expanding out of small scale factor
falls down its potential and recollapses with probability one,
in accordance with the classical solution;
but also, regions of the scalar potential corresponding
to a de~Sitter geometry give rise to string pair production,
which is seen in the large scale factor behavior
of the wavefunction.  
As we saw in section \ref{STinterp}, this pair production occurs
classically for tachyonic modes; it also occurs quantum
mechanically for massive modes, as we discuss in
the next subsection following the
analysis of \cite{Strominger:2002pc}.
It is interesting that the negative energy de~Sitter
solutions analogous to $J_\nu$ for real $\nu$
have disappeared from the sensible spectrum;
namely, if we consider \pref{nicefn} for real $\nu$,
the wavefunction has a nice oscillatory
behavior in the far future but necessarily blows up
in the far past

There is of course a second solution of the Schr\"odinger
equation satisfying the periodicity requirement:
\be\label{badfn}
  \Upsilon_\nu(z,\bar z)=\tilde C_\nu\Bigl[
	2\cos\nu\pi \;H_\nu^{\sst(1)}(z)H_\nu^{\sst(1)}(\bar z)
	-e^{-\nu\pi i}\Bigl(
		H_\nu^{\sst(1)}(z)H_\nu^{\sst(2)}(\bar z)
		+H_\nu^{\sst(2)}(z)H_\nu^{\sst(1)}(\bar z)
	\Bigr)\Bigr]
\ee
where again $\tilde C_\nu$ is a normalization.
These functions have the asymptotic behavior
\be\label{badass}
  \tilde C_\nu^{-1}\,\Upsilon_\nu(z,\bar z)\sim
	\cases{
	\frac{2i\,e^{-2\pi i\nu}}{\sin\,\pi\nu}
			[(z\bar z)^{\nu}+(z\bar z)^{-\nu}]
		&\qquad$|z|\to 0$ \cr
		& \cr
	\frac{4}{\pi|z|}\;\cosh[i(z-\bar z)]
		& \qquad $|z|\to\infty$\ .} 
\ee
These wavefunctions thus blow up as 
$\cosh[2\exp(\coeff\alpha2\phi)\sin(\coeff\alpha2 X)]$
for large scale (and also at small scale
for real $\nu$), and so have the same large scale asymptotics
as the de~Sitter wavefunctions
\pref{wdwsoln} when they are analytically continued
to the AdS sign of $\mu$.  We would like to interpret
the wavefunctions \pref{badass} similarly
as being related to spatially noncompact
solutions with AdS asymptotics. 
There are indeed spatially noncompact 
solutions of the complexified Liouville model,
which are just the complexification of the
static AdS geometries \pref{globalads}, \pref{poincare},
and \pref{rindlerads}.
We might choose to discard these wavefunctions
for the same reason, namely that the exponential growth
at large scale is related to static AdS geometries,
and not some `proliferation of de~Sitter space'.
Note again that the imposition of Feynman boundary
conditions in the far past will necessarily
involve these solutions.
We will return to a discussion of boundary conditions
in the Wheeler-DeWitt equation in section \ref{discsec}.


\subsection{\label{stringprodsec}String production}

In the semiclassical regime of large $d$, there are many
tachyons in the string spectrum.  They will all be generated
at the quantum level, and once generated, they will grow as
$\TT\sim\exp[(|m^2|-p^2)^\half t]$ for a mode of spatial
momentum $\vec p$.  If we regard the coefficient of
the exponent as a measure of the decay rate,
we can estimate the total decay rate per unit volume
from all the tachyons as
\bbb
  \Gamma_{\rm total} &=& 
	\int_0^{\sst\sqrt{{d}/{12}}}\! dm\int\! d^dp\,\rho(m)\sqrt{|m^2|-p^2}
\nonumber\\
	&\sim& {\rm const.}\times d^{\frac{d+1}{2}}\exp[4\pi d/12]\ ,
\label{lifetime}
\eee
where we have used the asymptotic density of states
$\rho(m)\sim \exp[4\pi\sqrt{d/12}\,m]$.
The result \pref{lifetime} can be regarded as an estimate
of the overall rate of growth of instabilities
in the target space background.

We saw in the classical solutions \pref{globalds}
and quantum minisuperspace wavefunctions \pref{wdwsoln}
that there is unsuppressed classical production of strings at
scale factors up to $e^{\gamma\phi}\sim 4/|\mu|$
(the minimum scale factor of global de~Sitter space).  
This is to be expected for the tachyonic modes 
of the string field, simply reflecting the instability of the target
space background.  However, pair production also occurs
for massive modes of the string field, as we now discuss.

Here, we are implicitly assuming the equivalence
of $2d$ cosmologies and closed strings, with
the naturally associated particle/string interpretation
of in and out states at $\phi\to\pm\infty$,
as \eg\ in \cite{Birrell:1982ix}.
Thus, for instance, a de~Sitter cosmology that
is usually thought of as contracting and then re-expanding,
is rather thought of as the creation of a 
`universe--anti-universe pair'; these differ only
in the relation between the direction
of the expansion of the scale factor and coordinate time.
Since what we physically detect is not coordinate time
but some thermodynamic arrow of time that agrees with
the direction of cosmic expansion, this is not
expected to be an observable distinction.
In other words, there is no `collapsing de~Sitter universe',
a de~Sitter universe propagating backwards in 
`scale factor time', rather there is a
$\overline{\it de~Sitter}$ universe%
\footnote{Unfortunately, we cannot call it
an anti-de~Sitter universe.}
propagating forward in scale factor time.

The pair production rate of massive open strings in 
an exponential tachyon background
was estimated in the minisuperspace approximation 
in \cite{Strominger:2002pc,Gutperle:2003xf}.
The closed string calculation is essentially the
same in the minisuperspace approximation, since the
zero mode feels the same exponential potential.
For de~Sitter space $\mu<0$, the pure incoming
plane wave solutions are the $J$ Bessel functions
\pref{wdwsoln}, whereas the pure outgoing plane wave
solutions $\Psi_{\omega}^{\it out}(\phi)$
are the Hankel functions \pref{Hbessel}.
The result for the appropriate ratio of Bogoliubov coefficients 
giving the transformation between in and out bases is%
\footnote{We thank A. Strominger and T. Takayanagi
for pointing out an error in this and the
following expression in an earlier version
of this paper.
Also, in order to facilitate comparison to 
\cite{Strominger:2002pc,Gutperle:2003xf}, 
we restore the factors of $\alpha'$ in this and the
following expressions.  In the rest of the text,
we have set $\alpha'=2$ in accordance with the conventions
of \cite{Ginsparg:1993is}.}
\be\label{prodrate}
  |\gamma_{\omega}|=\exp[-2\pi\omega\sqrt{\alpha'}]
\ee
(recall $\omega\sim m/\sqrt{2}\gamma$).
This ratio gives the probability amplitude to pair produce 
any individual heavy ($\omega\gg 1$) string mode.  
The typical string is produced at around $\phi=0$,
where the scale factor is of order one.

The pair production probability $|\gamma_{\omega}|^2$ is 
exponentially suppressed in the mass of the string state;
however, this is compensated by the exponentially large 
density of states at energy $\omega$,
\be\label{hagdens}
  \rho(\omega)\sim \exp\Bigl[+4\pi\omega\sqrt{\alpha'}
	\cdot \gamma\sqrt{\coeff{d-1}{12}}\Bigr]\ .
\ee
Since $\gamma\sqrt{\frac{d-1}{12}}>1$,%
\footnote{In the semiclassical limit, 
$\gamma\sqrt{\frac{d-1}{12}}\sim 1+\frac{6}{d-1}+...$,
and in the critical dimension, $\gamma\sqrt{\frac{d-1}{12}}=2$.}
the exponential suppression
of the Bogoliubov coefficients is always outweighed
by a faster growth in the density of states;
the total rate of production of massive string modes diverges.
Once produced, such strings gain energy 
exponentially rapidly from the
unbounded Liouville potential.  Their back reaction becomes
important at some finite time.  The overall
picture is of a Hagedorn gas of strings
being formed by the exponential time dependence of the
target space background.

For $\mu>0$, the tachyon condensate seems to have the 
opposite effect; classically, strings are `repelled from the future'.  
As mentioned above, the wavefunction that
describes an expanding and recollapsing AdS solution 
is exponentially damped at large scale factor --
there are no strings at late time.

The wavefunction \pref{nicefn} of complexified
Liouville theory exhibits the large scale factor
asymptotics of Hankel functions \pref{Hbessel} 
appropriate to pair production in de~Sitter space,
even though the reflection probability at small scale
factor is one.  We interpret this result as an indication
that the incoming cosmology follows its classical trajectory
and recollapses; yet at the same time, for any Liouville energy,
there is pair production
of strings from the region in $X$ where the matter potential
generates the de~Sitter sign of the cosmological constant.
This leads to a nonzero amplitude to find
a $2d$ universe at large scale factor.


\subsection{\label{thermosec}de Sitter thermodynamics}

While the statistical mechanics underlying the entropy
of gravitating systems in anti de~Sitter space has by now
achieved a rather firm footing 
(\cf\ \cite{Aharony:1999ti} for a review),
gravitational entropy in de~Sitter space has remained
a bit more mysterious.
There have been a number of analyses of
the entropy of two-dimensional de~Sitter space,
for example using ideas of quantum entanglement
\cite{Hawking:2000da}, or considering topological
gravity and its asymptotic symmetry algebra
\cite{Cadoni:2002kz,Medved:2002tq,Astorino:2002bj}
(following the ideas of \cite{Strominger:1998eq}).

One should be careful in drawing conclusions about
de~Sitter thermodynamics, and in comparing
different approaches to it, since the thermodynamic
quantities of energy, entropy, and temperature are
observer dependent.  For example, the entanglement
approach of \cite{Hawking:2000da} takes the point of
view of a static observer, considering the entanglement
of their observable degrees of freedom with those behind
their horizon; energy and temperature are those measured 
by the static observer.  On the other hand,
the approach of \cite{Cadoni:2002kz,Medved:2002tq,Astorino:2002bj}
uses asymptotic symmetries on the spacelike slice at
conformal infinity, and is thus necessarily observing the
entire state.  In particular, it is not clear how to 
relate the energies of the two approaches,
nor is it clear what interpretation
to give the entropy computed in 
\cite{Cadoni:2002kz,Medved:2002tq,Astorino:2002bj}.%
\footnote{It seems most closely related to the 
`observable entropy' of the $N$-bound \cite{Bousso:2000nf}
discussed below.}

In more than two dimensions, 
the de~Sitter gravitational entropy is the canonical
\be\label{BHent}
  S=\frac{A}{4G}\ ,
\ee
where $A$ is the area of the horizon
of a static observer, and $G$ is Newton's constant.
In two dimensions, this formula is rather ambiguous --
the horizon of the static patch is a zero-dimensional sphere,
which is only a pair of points; and 
one must identify the quantity playing the role of $G$, 
since there is no Einstein gravity per se.
The two approaches \cite{Hawking:2000da} and
\cite{Cadoni:2002kz,Medved:2002tq,Astorino:2002bj}
make differing assumptions, but both arrive in the end at 
expressions of the form
\be\label{twodent}
  S_{\rm 2d} = {\it const.} \times c
\ee
where the constant depends on the assumptions made,
and $c$ is the relevant conformal central charge of the system.
Note that we obtain this sort of answer if
we treat (as in \cite{Hawking:2000da}) both the
horizon area and the gravitational coupling
as dimensionless numbers.  The horizon area is 2,
for instance if we evaluate the general
formula ${\it Vol}(S^{n-1})=2\pi^{n/2}/\Gamma(\half n)$ for $n=1$,
or equivalently since the zero-dimensional transverse sphere 
consists of two points;
and in section \ref{LSGmodel}, we saw that
$8\pi G=\hf\alpha^2=6/c$.  These values yield 
\be\label{naivent}
  S_{\rm 2d}^{\sst\rm BH}=\frac{2\pi}{3}\, c
\ee
as the Bekenstein-Hawking
entropy of two-dimensional de~Sitter space.

Let us apply yet another heuristic argument for this form of the 
de~Sitter entropy.  A route to the de~Sitter entropy
in higher dimensions is through the Schwarzschild
de~Sitter geometry \cite{Spradlin:2001pw}.
In $n+1$ dimensions, the Schwarzschild de~Sitter geometry is
\bbb
  ds^2 &=&
	-\Bigl(1-\frac{r_0^{n-2}}{r^{n-2}}-\frac{r^2}{\ell^2}\Bigr)\, dt^2
	+\Bigl(1-\frac{r_0^{n-2}}{r^{n-2}}-\frac{r^2}{\ell^2}\Bigr)^{-1}dr^2
	+r^2 d\Omega_{n-1}^2
\nonumber\\
  r_0^{n-2} &=& \frac{16\pi G_{n+1}\,m}{(n-1){\it Vol}(S^{n-1})}\quad .
\label{sdsmetric}
\eee
Blithely continuing to $n=1$, after the shift 
$r\to r-\coeff{4\pi G_2 m\ell^2}{n-1}$
one finds 
\bbb
  ds^2 &=& -(a^2-r^2/\ell^2)\,dt^2+\frac{dr^2}{(a^2-r^2/\ell^2)}
\nonumber\\
  a^2 &=& [1+(\coeff{4\pi G_2\,m\ell}{n-1})^2]
\label{staticmetric}
\eee
with $\ell^2=-4/\mu$.  
This metric is appropriate to the static patch of an inertial
observer in the global geometry \pref{globalds},
where the time $t/\ell$ along that observer's worldline
in \pref{staticmetric} is identical
to the proper time in \pref{FRW}.  As a result,
we may identify $\epsilon=a$.
Demanding regularity of the Euclidean continuation 
$t_{\sst\rm E}=it$ of \pref{staticmetric}
requires $t_{\sst\rm E}\approx t_{\sst\rm E}+2\pi\ell/a$,
and determines the temperature $T=a/2\pi\ell$.
Note that one has $T\sim E$ at large $E$, 
appropriate to a $0+1$ dimensional gas.  
This suggests that there might be a model for the 
entropy in terms of a reservoir of quantum mechanical
degrees of freedom on the de~Sitter horizon.

For geometries that are asymptotically
de~Sitter in both past and future, $m^2<0$; so after continuing $M=im$
one has
\be
  dS_{\rm 2d} = -\frac{dE}{T}=
	- \frac{2\pi \ell\,dM}{\sqrt{1-(\coeff{4\pi G_2M\ell}{n-1})^2}}
\label{staticent}
\ee
(the sign is due to the fact that the energy under discussion
is being extracted from the horizon, not added to it).
The integral should run from the point of zero entropy $\epsilon=0$
to the value of interest; one finds
\be\label{enentrel}
  S_{\rm 2d}(M)= \frac{n-1}{2 G_2}\, 
	\cos^{-1}\Bigl(\frac{4\pi G_2M\ell}{n-1}\Bigr)\ .
\ee
One thus finds the entropy of global de~Sitter space 
by this reasoning to be
\be\label{goofy}
  S_{\rm dS,2d}=\frac{\pi(n-1)}{4 G_2}\ .
\ee
The factor $(n-1)/G_2$ is best understood from the
dimensional continuation of the Einstein action.
Consider the standard gravitational action 
$\coeff{1}{16\pi G_{n+1}}\int\!\!\sqrt{g}R$ in a metric which
is a Weyl rescaling $g=e^\rho \hat g$ of a fiducial
metric $\hat g$, near $n+1=2$ dimensions.  One finds
\bbb
  \frac{1}{16\pi G_{n+1}}\int\sqrt{g}R
	&=& \frac{1}{16\pi G_{n+1}}\int e^{\frac{n-1}2\rho}\sqrt{\hat g}
		\left[\hat R -n\hat\nabla^2\rho-\coeff{n(n-1)}{4}
			(\hat\nabla\rho)^2\right]
\nonumber\\
	&{\buildrel {n\sim1} \over {\longrightarrow} }&
	 \frac{1}{16\pi G_{n+1}}\int \sqrt{\hat g}
	\left[\hat R+\coeff{n-1}{4}(\hat\nabla\rho)^2
		+\coeff{n-1}2\rho\hat R+\ldots\right]\ .\qquad
\label{neartwo}
\eee
Thus the Liouville action arises as the leading non-topological
term in the Einstein action expanded around two dimensions.
Comparing to \pref{Sliou}, we identify
\be\label{einnewt}
  \frac{n-1}{G_2} = \frac{8}{\gamma^2} = \frac{2c}{3}\ .
\ee
Plugging into \pref{goofy}, we find
\be\label{dubious}
  S_{\rm dS,2d}=\frac{\pi}{6}\,c
\ee
which is $1/4$ of the answer \pref{naivent}.
But we should be suspicious of the naive Bekenstein-Hawking
answer \pref{naivent}; it suggests that the entropy is
independent of the energy, and thus does not decrease
if we add some particles to the de~Sitter vacuum.
On the other hand, it is reasonable to expect that
$dS=-dE/T$; extracting particles from the de~Sitter
horizon is only consistent with constant $S$ 
for infinite Hawking temperature,
which is incompatible with the finite 
periodicity of the Euclidean continuation of the geometry.

One might question some of the assumptions made
in the derivations that lead to \pref{naivent}
and \pref{dubious}, for instance the evaluation
of the Newton constant is different in the two cases.
Also, in the analysis of the first law, we dimensionally 
continued away from two dimensions $n=1$.
Let us do the same for the area law.
Consider the cosmological horizon of the
Schwarzschild de~Sitter solution \pref{sdsmetric};
as $n\to1$, it is located at
\be
\label{horrad}
  \frac{r_{\rm hor}}{\ell} = -\,\frac{\ell}{2r_0}+
	\sqrt{\Bigl(\frac{\ell}{2r_0}\Bigr)^2+1}\quad ,
\ee
where
\be
\label{rzerotwod}
  \frac{\ell}{r_0} = \frac{8\pi G_2m\ell}{n-1}\quad,
\ee
The $S^{n-1}$ horizon sphere 
has area ${\it Vol}(S^{n-1})\;(r_{\rm hor})^{n-1}$,
so as $n\to1$
\be\label{areavar}
  dA = 2(n-1)\,d\log\left[-\frac{4\pi G_2m\ell}{n-1}
	+\sqrt{\Bigl(\frac{4\pi G_2m\ell}{n-1}\Bigr)^2+1}\;\right]
\ee
Again continuing $M=im$, we find
\be
\label{diffarea}
  \frac{dA}{4G_2} = \frac{n-1}{2G_2}\; 
	d\cos^{-1}\Bigl(\frac{4\pi G_2M\ell}{n-1}\Bigr)\quad ,
\ee
in agreement with \pref{enentrel}.  Thus, the
dimensional continuation of the area law
agrees with the corresponding continuation of
the first law.

Finally, it has been argued in \cite{Bousso:2000nf}
that the entropy of de~Sitter space should
bound the `observable' entropy of matter in a space that is
asymptotically de~Sitter in both the past and future.
The basic idea is that extracting too much entropy from
the de~Sitter horizon of a static observer
requires putting too much energy into the spacetime,
and hence a cosmological singularity develops either in
the past or future.  Let us see how this argument
works in two dimensions.

Global de~Sitter space is the vacuum $\epsilon=1$ of the 
solution \pref{globalds}, where the matter fields
are also unexcited.  As we populate the levels of the matter
system, $\epsilon$ decreases due to the constraint
\pref{enconstraint}, until it reaches zero
at $E_{\rm matter}=c/24$. 
At this point the geometry is on the cusp
of having a singularity, and indeed there will be a big crunch
(Milne singularity) if we add more matter energy, since the 
solutions will cross over into the hyperbolic class 
\pref{milneds}.  Note that the addition of matter
makes the de~Sitter space `taller', see 
figure \dsdomains; 
higher-dimensional gravity also exhibits this
property, as shown in \cite{Gao:2000ga}
and applied to holographic bounds in
\cite{Bousso:2002ju,Bousso:2002fq}.

Due to the exponential growth
in the density of levels, the total number of matter states in
the class of geometries which is asymptotically de~Sitter
both in the past and in the future is bounded by
the density of levels at $L_0=\bar L_0=c/24$; 
in this way, one reaches an analogue of the 
`$N$-bound' of \cite{Bousso:2000nf}
\be
  S_{\rm matter} =
	2\pi\sqrt{\coeff c6 L_0}+2\pi\sqrt{\coeff c6 \bar L_0}
	\;\le\; \frac{\pi}{3}\,c = 2S_{\rm dS,2d}\ .
\label{globalent}
\ee
One sees from the previous considerations
that twice the de~Sitter entropy given by \pref{dubious} 
bounds the `observable' entropy of matter.
The bound argued for in 
\cite{Bousso:2000nf,Bousso:2002ju,Bousso:2002fq}
is actually $S_{\rm matter}\le S_{\rm dS}$
rather than twice $S_{\rm dS}$.  
It would be interesting to understand the source of
this discrepancy, and more generally, 
to understand why the entropy
\pref{enentrel} differs in form from \pref{globalent}.

The above results point the way toward an understanding
of de~Sitter quantum gravity and its thermodynamics.  
Global $2d$ de~Sitter space 
is the vacuum of matter coupled to 
the ground state of Liouville theory, and is thus
a unique state.%
\footnote{We are of course assuming that quantum Liouville
theory with this sign of the cosmological constant exists
as a well-defined Lorentzian CFT.}
One might then view the de~Sitter entropy of a static observer
as entirely an effect of quantum entanglement. 
Alternatively, global de~Sitter space only lives for
a finite amount of conformal time; one might also explain the
entropy as an inability to decide precisely the number of
quanta in the global CFT state, since field modes
do not undergo a full period of oscillation in the
time available to observe the state.  However, de~Sitter
entropy seems rather different from black hole entropy, 
which for instance in the AdS/CFT correspondence 
is a count of distinct microstates of quantum gravity.


\section{\label{discsec}Discussion}


\subsection{\label{eternalchaos}The beginning of the beginning}

One of the outstanding issues facing inflationary cosmology is the
origin of the inflationary phase.  A number of suggestions
have been put forward in this regard:
\begin{itemize}
\item
The `old' \cite{Guth:1981zm}
and `new' \cite{Linde:1982mu,Albrecht:1982wi}
inflationary scenarios, whereby the universe is
assumed to start off in a 
metastable or slightly unstable inflationary phase;
\item
The various `no-boundary' or `tunneling' proposals 
\cite{%
Hawking:1982fz,%
Hartle:1983ai,%
Vilenkin:1983xq,%
Linde:1984mx,%
Vilenkin:1986cy,%
Vilenkin:1988kf,%
Hawking:1998bn%
}
whereby the initial inflationary
state is determined by a tunnelling event;
\item
`Chaotic' inflation 
\cite{Linde:1983gd},
in which the initial value of
the inflaton field is taken to be a random variable,
so that some fraction of the ensemble of initial
conditions leads to inflation;
\item
`Eternal/chaotic' inflation
\cite{Vilenkin:1983xq,Steinhardt:1982kg,Linde:1986fd,Linde:1994xx},
whereby quantum fluctuations of the inflaton
drive the volume-weighted average of the field to 
climb its potential.  The `typical' FRW domain is
surrounded by, and came from, an inflationary phase.
\end{itemize}

Each of these scenarios has its drawbacks.  Old inflation 
does not end gracefully, producing unacceptably large
fluctuations; new inflation begins by fiat with a
specially tuned initial condition -- a quiescent scalar
field at a very flat local maximum of its potential -- 
and so to some extent begs the question of how the universe
became large, flat, and isotropic.
One must find some measure on the space of initial
conditions for cosmology, explain why the inflating
solutions have a non-negligible probability, and so on
(\cf\ \cite{Hollands:2002xi} for a recent discussion).

The no-boundary/tunnelling proposals are an attempt to provide such
a measure which makes the inflating state a reasonable
starting point.  However, one could also regard it
as an arbitrary selection criterion on states,
based on a notion of simplicity -- typically only `nice'
(simple, relatively symmetric)
states have a `nice' (nonsingular) Euclidean continuation.
In quantum field theory any nontrivial S-matrix process
does not have such a `nice' Euclidean continuation;
there, the proposal would throw out essentially 
all the interesting structure!
In the context of $2d$ models, any state (other than
the CFT vacuum) with a truly nonsingular
Euclidean continuation would amount to a tadpole
amplitude for the string background, which would
then not satisfy the classical string equations
of motion.

Chaotic inflation is to some extent a similar selection
criterion.  The inflaton field is assumed to lie in a chaotic
distribution of initial conditions, or in other words its
quantum wavefunction is broadly distributed.
However the wavefunction should not be so chaotic that
the inflaton kinetic energy dominates over its potential
energy; at least, some regions of space should be sufficiently
quiescent for a sufficiently long period that 
sufficient inflation takes place.  For instance, a hot big bang
is not the right kind of initial chaos if the potential is
$V(X)=m^2X^2$.  Nevertheless, there will be a relatively
large phase space of acceptable initial conditions.
Note that inflating wavefunctions
of our two-dimensional models have somewhat the character
of chaotic inflation -- the quantum wavefunction
can be taken to be more or less uniformly smeared over all values of
the inflaton field $X$ (quantum fluctuations in two dimensions
do this naturally).  Some fraction of the distribution
will inflate along trajectories of the sort described in 
section \ref{LSGmodel}.

In the context of minisuperspace models, the discussion
of appropriate boundary conditions to be imposed on the 
Wheeler-DeWitt wavefunction has a rich and varied history
(for reviews, see for instance 
\cite{Vilenkin:1994rn,Wiltshire:1995vk}).
In pure Liouville gravity, the minisuperspace
`no-boundary' proposal \cite{Hartle:1983ai}
amounts to the selection of wavefunctions $\Psi=J_\nu$ for real $\nu$
(whose classical geometries are the global
de~Sitter geometries \pref{globalds}).
These real-valued functions
behave as the cosine of the scale factor $\cos(\aa)$
at large scales, and as a power law of the scale which vanishes 
as $\aa^\nu$ at small scales.  The tunneling proposal
of \cite{Vilenkin:1986cy} selects purely expanding
universes at large scale, which are the Hankel functions
$H^{\sst(2)}_\nu$.  Yet another choice
suggested in \cite{Wald:1993kj} is to 
consider only expanding universes at small scale,
which selects $\Psi=J_\nu$ for imaginary $\nu$.
The latter is particularly related to the implementation
of Feynman boundary conditions and the associated
interpretation of $2d$ cosmologies as quanta
in a string field theory.  This interpretation
arose some time ago in the context of minisuperspace
models of baby universes
(\cf\ \cite{Strominger:1988ys} for a review,
and also \cite{Rubakov:1988jf,McGuigan:1988vi,McGuigan:1989es,%
Hosoya:1989aa,Fischler:1989ka}).

Finally, let us turn to the chaotic/eternal inflationary
scenario.  Its logic rests 
on a number of assumptions that are open to question.
First of all, given the apparent success of the notion of
holography in black hole physics, one might wonder whether
the appropriate probability measure of a geometry
in quantum gravity is related to its proper volume,
rather than for instance the area of its horizon
or some other measure.%
\footnote{Replacing volume by area would not however 
seriously affect the estimates of chaotic inflation \eg\ in
\cite{Linde:1994xx}.}
Second, as in black hole physics
it is dangerous to ascribe objective reality
to properties of spacetime that lie beyond the horizon of
any observer in the spacetime.  Any given observer sees a
finite probability per unit proper time for inflation to terminate,
thus there is no observable that detects an eternal proliferation
of inflating domains.%
\footnote{At least, no local observable.  There might
be nonlocal observables; they would have to be stitched together
out of holographic data on the horizons of all local
observers, taking proper account of the entanglement of this data 
(\cf\ \cite{CarneirodaCunha:2001jf,Maldacena:2001kr,banksidea}).}
Third, the notion of eternal inflation
has a Zeno-like quality~-- one takes the tiny fraction of
the probability density where the field jumps up the potential,
counter to the dictates of classical dynamics; that region
wins the competition of volume growth as the field slowly rolls
back down; then one takes the further tiny fraction of the probability
density where the field jumps up the potential; and so on.
It would seem that at some point one will run out of
probability density.  
Fourth, the domains of large proper volume so 
obtained occupy a vanishingly small (conformal) coordinate domain,
which is ostensibly compensated by a huge conformal factor in
that region, leading to the typical fractal conformal diagram of the
late-time structure of eternal/chaotic inflation
(see figure \fractal). 
But how are we to distinguish these violent fluctuations 
of the conformal factor from the standard fluctuations
of a quantum field on all scales, that we are accustomed
to regularizing and renormalizing away?

The essential driving force behind this violent oscillation
of the scale factor is the negative metric
on the kinetic energy of the scale factor.
In the context of two-dimensional models of the sort 
we have been considering, many if not all
of the fluctuations of the scale factor (Liouville mode)
are gauge artifacts, which are eliminated by the
BRST constraints on the Wheeler-DeWitt wavefunction.
The physical state space has positive metric.
In fact, in the initial state at $\phi\to-\infty$,
the matter potential turns off and all the fields
become free.  Then, up to BRST trivial shifts
$\ket{\Psi}\to \ket{\Psi}+Q_{\sst\rm BRST}\ket{\Upsilon}$,
we can consider states having only the zero mode
of the scale factor excited.  It is hard to see how
a fractal structure of the scale factor is going to emerge,
given that the Virasoro constraints dictate
the future evolution of the scale factor in terms
of the state of the matter fields; it would seem that
such violent fluctuations would have to be built in
by the choice of particular highly excited matter states,
which is hardly the appropriate starting point for inflation.

\begin{figure}[ht]
\begin{center}
\[
\mbox{\begin{picture}(169,204)(0,0)
\includegraphics{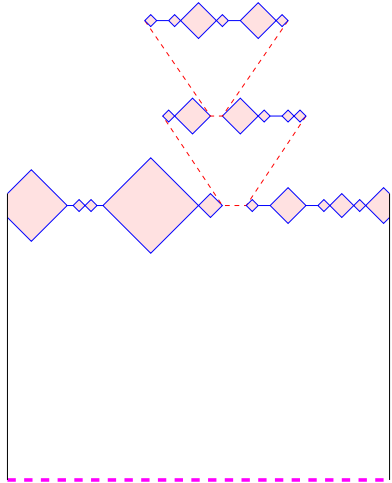}
\end{picture}}
\]
\caption{\it
The fractal structure of the eternally/chaotically
inflating universe.  The shaded domains represent
the ordinary FRW cosmological regions that arise when
the chaotic inflaton sporadically falls out of
the inflationary regime.
In a two-dimensional model, such configurations
represent violent UV fluctutations of the Liouville mode,
which might not have physical significance.
}
\end{center}
\end{figure} 

One way that the notion of chaotic inflation might be consistent
with the analysis of the preceding sections would be if
the fluctuations of the scale factor had an interpretation
in terms of universe production.  We saw an example of this
in the context of topological inflation, wherein
a singularity in the scale factor found an interpretation in
terms of the branching off of a child universe that carried
off the core of the inflating domain wall.  
Perhaps when one is able to calculate reliably,
the fluctuations of the scale factor in chaotic inflation
will be seen to actually reflect the creation of child
universes.  However, note that in the minisuperspace
approximation, the appearance in the wavefunction
of strings having large scale factor occurred by 
mechanism separate from the dynamics of the expanding
universe at early time; that universe returned
to zero scale factor with probability one,
and the universes appearing at late time were the
result of pair creation processes.


\subsection{\label{altscen}An alternative:
associated production of universes}

The string dynamics that results from tachyon condensation
has an amusing $2d$ cosmological interpretation, that
provides a new explanation for the origin 
of inflation.  Back reaction of the target space geometry
eventually becomes important, and determines the
actual late time (large scale factor) behavior
of the theory.  The late-time behavior of bulk tachyon
condensation is not well understood, but it is conceivable
that it will have the effect of saturating
the exponential growth of the worldsheet potential,
and thus relaxing the $2d$ cosmological constant.
Moreover, strings are pair produced by the condensate
and eventually become important in determining
the late time (large scale factor) behavior.
The production of large numbers of strings
($2d$ cosmologies) in association with a
tachyon condensate, which becomes an ordinary
supergraviton wave at late times, provides a 
new perspective on the origin of inflation.

Localized tachyon condensation provides a controlled
laboratory where precisely these mechanisms occur.
As we argued
in section \ref{localtach}, the late time behavior of the
condensate in the throat models is a pulse of radiation
travelling up the throat, leaving behind the throat of $k$
fivebranes.  The exponential growth in time of the tachyon $\TT$
saturates due to backreaction and evolves into an ordinary
sigma model background (albeit time-dependent).  
At late time, the worldsheet of a given string left behind 
by the outgoing pulse sees a flat potential, and the classical
solution of the Liouville field will be approximately
linear in worldsheet conformal time $\tau$,
$\phi\sim\varepsilon\tau$, corresponding to an expanding 
Milne geometry.  This is an FRW cosmology with scale factor
$a(t)\sim t$ -- the cosmological constant relaxes to zero
at late time.  Strings initially trapped in AdS regions
(large $|\partial_\phi\TT|$, small $|\partial_{\sst X}\TT|$) 
do not reach large scale factor $\phi$;
strings in the initial state finding themselves in an inflating region
(large $|\partial_{\sst X}\TT|$, small $|\partial_\phi\TT|$) 
do reach large scale factor,
and in addition there will be string pair creation by the time-dependent
background in these de~Sitter regions.
The latter effect generates a probability distribution on
$2d$ cosmologies.

We can then ask the question: 
What is the character of the typical $2d$ universe
at large scale factor $\phi$?  
The answer will be that it
is an expanding, radiation dominated FRW universe!%
\footnote{In the $2d$ energy constraint \pref{friedmann},
the energy density evolves as $\rho\sim{\rm const.}\times \aa^{-2}$,
which is the same as the `spatial curvature' term
$-(\coeff18 Q^2+1)\aa^{-2}$.  Thus, once radiation dominated,
always radiation dominated -- there will never be a critical
scale factor at which the latter term takes over and causes
recollapse.}
The estimates of \cite{Strominger:2002pc}
quoted in section \ref{stringprodsec}
indicate that the pair production rate is substantial;
there is of course also the `classical' production of
the tachyon modes themselves, which become the coherent
pulse of supergravitons propagating to infinity.
The latter is essentially the analogue of the exponential production
of (cold, empty) `baby universes' observed in 
\cite{Rubakov:1988jf,McGuigan:1988vi,McGuigan:1989es,%
Hosoya:1989aa,Fischler:1989ka}.
There are also many strings produced in excited states,
as we saw in section \ref{stringprodsec},
and it will be an important task to determine 
their distribution.
If the typical $2d$ universe left behind by the pulse
arises from such a pair production process,
we have a situation where observers living in
the typical late-time universe would see 
the structure of the sort we want to arise from inflationary theory,
namely that their universe experienced an early period of inflation 
which exited into an expanding FRW phase.
The problem of how inflation got started is neatly solved
by the fact that most of the (non-empty) $2d$ universes at late time
were created by a quantum process,
and the attempt to trace their geometries backward classically
to a big bang at $\phi=-\infty$ is not a valid procedure.
Note that this mechanism is distinct from the 
`tunnelling/no-boundary' proposals of
\cite{%
Hawking:1982fz,%
Hartle:1983ai,%
Vilenkin:1983xq,%
Linde:1984mx,%
Vilenkin:1986cy,%
Vilenkin:1988kf,%
Hawking:1998bn,%
Hawking:2002af}
and others.  There, the entire interpretation
of the wavefunction is in terms of
single universes.%
\footnote{Which appear from `nothing'
by tunnelling; however, as mentioned above,
the embedding of this idea in $2d$
models takes on the character of an
aesthetic criterion for the selection 
of an `in' state of the CFT.}
Here, the mechanism is the quantum mechanical 
Bogoliubov transformation between `in' and `out'
states of the string field.  Such effects will be present
even if the string coupling is tuned to be small.

Sadly, these late-time cosmologies have unrealistic
aspects (besides the fact that they are two-dimensional).
Being strings, they will always have moduli (that locate the
string in the smooth target space), and the matter energy
density of the late-time cosmology is radiation dominated,
residing in the fluctuations of these moduli.  
Intrinsic to our model is the problem 
of moduli production at the end of inflation.%
\footnote{And of course we may try to search for solutions.
For example, we can try to limit the number of flat,
runaway directions in the potential to one by constructing
higher-dimensional throats of the sort in \cite{Giveon:1999zm},
which have fewer noncompact `runaway' directions.}

It would certainly be interesting if such a scenario
could be exhibited in more realistic models.
There are certain features of string theory that parallel
what we have found here.  Known constructions of 
de~Sitter space in string theory
\cite{Maloney:2002rr,Kachru:2003aw} have the property
that the de~Sitter phase has finite duration, eventually
relaxing to an FRW phase (\eg\ as described in \cite{Craps:2002ii}
for the models of \cite{Maloney:2002rr}).  Thus 
the asymptotic geometry at large scale factor is 
always FRW.  Related to this, the qualitative structure
of the configuration space of string theory is the same
as in our two-dimensional model of section \ref{localtach} --
the `tachyon' (the cosmological constant)
is `localized' in field space; there always seem to
be one or more `flat directions to infinity' \cite{Dine:1985he},
with the nontrivial potential only in some finite region
of the low-energy field space.
In other words, the instability to exponential growth of 
the scale factor only exists in a localized region
of field space, and persists only for finite time,
much as in the $2d$ model.
Topology change is expected to occur in string theory
(\cite{Candelas:1989di,Aspinwall:1994nu,Strominger:1995cz,Witten:1998zw},
to name a few);
could topology change result in `universal pair production'?
Does this idea explain why the typical universe at large scale
emerged from an inflationary phase?%
\footnote{If so, one might imagine generalizations in various
directions.  For example, the number of macroscopic dimensions
of spacetime might be explained as a consequence of 
the genericity of an effective potential in four as opposed
to other dimensions, together with considerations of the
available phase space for universe creation.}

\vskip 2cm
\noindent
{{{\bf Acknowledgments}}}:
We thank
F. Larsen,
W. McElgin,
R. Wald,
and especially
D. Kutasov
for discussions.
This work was supported by DOE grant DE-FG02-90ER-40560.

\newpage


\providecommand{\href}[2]{#2}\begingroup\raggedright\endgroup

\end{document}